\documentclass[pdflatex,sn-mathphys-num]{sn-jnl}

\usepackage[utf8]{inputenc}
\usepackage[title]{appendix}
\usepackage[breakable]{tcolorbox}
\usepackage[numbers]{natbib}
\usepackage[nounderscore,square]{syntax}

\usepackage{adjustbox}
\usepackage{algorithm}
\usepackage{algorithmic} 
\usepackage{amsmath,amssymb,amsfonts}
\usepackage{amsthm}
\usepackage{array}
\usepackage{backnaur}
\usepackage{balance}
\usepackage{booktabs}
\usepackage{caption}
\usepackage{color}
\usepackage{csquotes}
\usepackage{diagbox}
\usepackage{dirtytalk}
\usepackage{etoolbox}
\usepackage{fdsymbol}
\usepackage{float}
\usepackage{graphicx}
\usepackage{hyperref}
\usepackage{lastpage}
\usepackage{listings,lstautogobble}
\usepackage{makecell}
\usepackage{manyfoot}
\usepackage{mathrsfs}
\usepackage{multirow}
\usepackage{ragged2e}
\usepackage{setspace}
\usepackage{subfig}
\usepackage{tabularx}
\usepackage{textcomp}
\usepackage{tikz}
\usepackage{tkz-graph}
\usepackage{url}
\usepackage{verbatim}
\usepackage[left, switch]{lineno}
\usepackage[varqu]{zi4} 

\usetikzlibrary{matrix,fit,arrows,calc,positioning,fadings,shapes,shadows,arrows.meta}
\definecolor{orcidlogo}{rgb}{0.37,0.48,0.13}
\definecolor{unilogo}{rgb}{0.16, 0.26, 0.58}
\definecolor{maillogo}{rgb}{0.58, 0.16, 0.26}
\definecolor{darkblue}{rgb}{0.0,0.0,0.0}
\definecolor{gray}{rgb}{0.4,0.4,0.4}
\definecolor{cyan}{rgb}{0.0,0.6,0.6}
\definecolor{codegreen}{rgb}{0,0.6,0}
\definecolor{codegray}{rgb}{0.5,0.5,0.5}
\definecolor{codepurple}{rgb}{0.58,0,0.82}
\definecolor{backcolour}{rgb}{0.95,0.95,0.92}
\definecolor{eclipseBlue}{RGB}{42,0.0,255}
\definecolor{eclipseGreen}{RGB}{63,127,95}
\definecolor{eclipsePurple}{RGB}{127,0,85}
\hypersetup{colorlinks,breaklinks,
	linkcolor=darkblue,urlcolor=darkblue,
	anchorcolor=darkblue,citecolor=darkblue}

\newcolumntype{L}[1]{>{\raggedright\let\newline\\\arraybackslash\hspace{0pt}}m{#1}}
\newcolumntype{C}[1]{>{\centering\let\newline\\\arraybackslash\hspace{0pt}}m{#1}}
\newcolumntype{R}[1]{>{\raggedleft\let\newline\\\arraybackslash\hspace{0pt}}m{#1}}
\newcolumntype{Y}{>{\centering\arraybackslash}X}
\newcommand{\algrule}[1][.2pt]{\par\vskip.5\baselineskip\hrule height #1\par\vskip.5\baselineskip}

\definecolor{gray}{rgb}{0.4,0.4,0.4}
\definecolor{darkblue}{rgb}{0.0,0.0,0.6}
\definecolor{cyan}{rgb}{0.0,0.6,0.6}
\definecolor{pblue}{rgb}{0.13,0.13,1}
\definecolor{pgreen}{rgb}{0,0.5,0}
\definecolor{pred}{rgb}{0.9,0,0}
\definecolor{pgrey}{rgb}{0.46,0.45,0.48}
\definecolor{codegreen}{rgb}{0,0.6,0}
\definecolor{codegray}{rgb}{0.5,0.5,0.5}
\definecolor{codepurple}{rgb}{0.58,0,0.82}
\definecolor{backcolour}{rgb}{0.95,0.95,0.92}

\lstdefinestyle{sas}{
	commentstyle=\color{codegreen},
	frame=single,
	language=java,
	xleftmargin=2.5em,
	xrightmargin=0.7em,
	framexleftmargin=1.5em,
	keywordstyle=\bfseries \color{black},
	numberstyle=\tiny\color{codegray},
	stringstyle=\color{codepurple},
	basicstyle=\ttfamily\tiny, 
	columns=fullflexible, keepspaces=true,
	breaklines=true, showstringspaces=false
	breakatwhitespace=false,       
	alsoletter=-,
	breaklines=true,                 
	captionpos=b,                    
	keepspaces=true,                 
	numbers=left,                    
	numbersep=5pt,                  
	showspaces=false,                
	showstringspaces=false,
	showtabs=false,                  
	tabsize=2,
	morekeywords={Architecture, Managing, Managed, LoopManager, Loop, , loop, Monitor, Analyzer, Planner, subSystem, inSubSystem, Component, Executor,Knowledge,Sensor,Effector,MeasuredOutput, Rules, monitor, analyzer, planner, executor, sensor,reference-input, must-use, must-not-use, effector, knowledge, ReferenceInput, measured-output,Alternative,withDomainRules }
}

\lstdefinelanguage{xtend}
{
	morekeywords={
		FOR,
		ENDFOR,
		structureElement,
		for
	},
	basicstyle=\ttfamily\tiny, 
	columns=fullflexible, keepspaces=true,
	breaklines=true, showstringspaces=false
	sensitive=true, 
	morecomment=[l]{//}, 
	morecomment=[s]{/*}{*/}, 
	morestring=[b]" 
}

\lstdefinelanguage{xml}
{
	morestring=[b]",
	morestring=[s]{>}{<},
	morecomment=[s]{<?}{?>},
	basicstyle=\ttfamily\tiny 
	columns=fullflexible, 
	keepspaces=true,
	breaklines=true, 
	showstringspaces=false
	numberstyle=\tiny\color{codegray},
	stringstyle=\color{black},
	identifierstyle=\color{darkblue},
	keywordstyle=\color{cyan},
	morekeywords={xmlns,version,type},
	postbreak=\mbox{\textcolor{red}{$\hookrightarrow$}\space},
	numbers=left,
	xleftmargin=0.7cm,
	frame=single
}

\lstdefinelanguage{QVT}{
	morekeywords={
		checkonly, domain, enforce, extends, implementedby, import, 
		key, overrides, primitive, query, relation, top, 
		transformation, when, where
		Bag, Collection, Dict, OrderedSet, Sequence, Set, Tuple, abstract, 
		access, and, any, assert, blackbox, break, case, class, collect, 
		collectNested, collectOne, collectselect, collectselectOne, 
		composes, compute, configuration, constructor, continue, datatype, 
		default, derived, disjuncts, do, elif, else, end, endif, 
		enum, except, exists, extends, exception, false, forAll, forEach , 
		forOne, from, helper, if, implies, import , in, inherits, init, 
		inout, intermediate, invresolve, invresolveIn, invresolveone, 
		invresolveoneIn , isUnique, iterate, late, let, library, literal, 
		log, main, map, mapping, merges, metamodel, modeltype, new, not, 
		null, object, one, or, ordered, out, package, population, primitive, property, 
		query, raise, readonly, references, refines, reject, resolve, resolveIn, resolveone, 
		resolveoneIn, return, select, selectOne, sortedBy, static, 
		switch, tag, then, transformation, true, try, typedef, unlimited, 
		uses, var, when, where, while, with, xcollect , xmap, xor, xselect 
	},
	keywordstyle=[2]{\textbf},
	morecomment=[l]{--},
	morestring=[b]{'},
	basicstyle=\ttfamily\tiny, 
	columns=fullflexible, keepspaces=true,
	breaklines=true, showstringspaces=false,
	tabsize=4}

\begin{document}

\title[Architectural Conformance Checking for MAPE-K-based Self-Adaptive Systems]{Architectural Conformance Checking for MAPE-K-based Self-Adaptive Systems}


\author*[1]{\fnm{Daniel} \sur{San Martín}}\email{daniel.sanmartin@ucn.cl}
\author[2]{\fnm{Valter} \sur{Vieira de Camargo}}\email{valtervcamargo@ufscar.br}
\author[2]{\fnm{Guisella} \sur{Angulo}}\email{guisella.angulo@estudante.ufscar.br}

\affil*[1]{\orgdiv{Escuela de Ingeniería}, \orgname{Universidad Católica del Norte}, \orgaddress{\city{Coquimbo}, \country{Chile}}}
\affil[2]{\orgdiv{Departamento de Computação}, \orgname{Universidade Federal de São Carlos}, \orgaddress{\city{São Carlos}, \state{SP}, \country{Brazil}}}


\abstract{Self-adaptive systems (SASs) adjust their behavior at runtime in response to internal or external change. The MAPE-K model, which includes Monitors, Analyzers, Planners, Executors, and shared Knowledge, is a reference for structuring feedback loops. As SASs evolve, implementations can drift from the intended MAPE-K architecture, compromising planned quality attributes. Architectural Conformance Checking (ACC) addresses this risk by comparing the current implementation to a specification of the architecture. General purpose ACC techniques are flexible, but lack SAS specific semantics, leading to ambiguous specifications and missed violations. We present REMEDY, an ACC approach designed for MAPE-K based SASs. REMEDY provides three elements: a domain specific language for expressing planned architectures in MAPE-K terms, a tool that extracts the implemented architecture, and a conformance engine that reports violations. By encoding SAS domain rules and reusing MAPE-K abstractions, REMEDY reduces specification effort and lowers error rates relative to general ACC. We evaluate REMEDY through a robotic SAS case study and a controlled experiment with software engineering students. Results show higher modeling productivity and effective detection of architectural drift, supporting more reliable verification of conformance to the MAPE-K reference model.
}

\keywords{self-adaptive systems, architectural conformance checking, architectural drift}
\maketitle

\section{Introduction}

Self-Adaptive Systems (SASs) are systems able to adapt themselves to internal and external changes. They are increasingly useful in domains where software must autonomously react to unpredictable changes, ranging from healthcare and finance to robotics and smart infrastructure. To perform adaptations, these systems continuously monitor operational conditions and adjust their behavior in response, reducing the need for human intervention. To support this adaptive behavior, they are often designed considering the concept of control loops for sensing, reasoning, and acting upon changes. 

MAPE-K is a reference architectural model for  designing SAS. It prescribes the main abstractions a SAS must exibit, as well as how these abstractions must communicate, i.e., their dependencies or communication rules. The main abstractions prescribed are Monitors, Analyzers, Planners, Executors and Knowledge base. The consequence of structuring a SAS with MAPE-K is a clearer architecture and better maintainability and evolution levels 
 \cite{Ibm2005, Brun2009, WeynsBook2021}. 
 
In general, the architecture of a SAS is well aligned with MAPE-K in the first releases, when only a few maintenance activities have been performed. At this stage, the MAPE-K abstractions are still clearly mapped to the source code, and the communication rules are still respected. However, as time passes and many maintenance activities are performed without sufficient rigor, the system’s architecture begins to deviate from the MAPE-K structure. This occurs when modifications cause the source code of an abstraction to spread across different parts of the system, and/or when communication rules are no longer respected.
These inconsistencies, known as architectural drift, manifest as violations of structural or behavioral constraints and can lead to increased technical debt, higher maintenance effort, and reduced system clarity \cite{Krupitzer2015, Weyns2012, Weyns2019}. Detecting such deviations is particularly important in SASs, where the adaptation logic is deeply intertwined with the system architecture.  

Architectural Conformance Checking (ACC) provides a solution to this challenge by comparing the Current Architecture (CA) of a system against a predefined Planned Architecture (PA) \cite{Pruijt2017}. The typical ACC process involves three main stages: (i) defining the PA using a modeling language or metamodel that specifies the required architectural abstractions and communication rules; (ii) recovering the CA by mapping implementation-level artifacts (e.g., classes, modules, interfaces) to these abstractions; and (iii) checking for conformance between PA and CA, identifying any mismatches. 

Most ACC techniques found in literature are domain-independent, that is, they can be used for any system, regardless of the domain the system belongs to \cite{Passos2010, Rosik2011, Pruijt2013, Bucaioni2023, Singjai2022, Mendoza2021}. While acknowledging the benefits of generic approaches, they also exhibit some deficiencies like i) the using of generic architectural terms for specifying domain-specific abstractions, leading to too general specifications and; ii) the need of specifying abstractions and communication rules that could be avoided when using a domain-specific approach. These can introduce errors, increase the effort, and result in inconsistent specifications. Domain-specific methods can alleviate this burden by embedding known patterns and rules into the modeling and verification process \cite{Krieger2020, Oliveira2022, Lima2020}.

To structure our investigation, we pose the following research questions:

\begin{itemize}
  \item \textbf{RQ1:} How can the definition of planned architectures for MAPE-K-based SASs be improved using a domain-specific language?
 
  \item \textbf{RQ2:} Can an automated conformance checking process effectively identify architectural drift in real SAS scenarios?

  \item \textbf{RQ3:} Does the use of a MAPE-K-specific ACC method reduce modeling effort and improve precision compared to a general-purpose DSL?
\end{itemize}

To address these limitations, we propose REMEDY, an ACC method specifically designed for SASs built upon the MAPE-K model. REMEDY incorporates three main elements: (i) a Domain-Specific Language (DSL-REMEDY) that allows software architects to define planned architectures using MAPE-K abstractions; (ii) an automated recovery tool that maps implementation-level artifacts to MAPE-K elements; and (iii) a conformance checking engine that detects and visualizes architectural drifts. By relying on a metamodel grounded in MAPE-K, REMEDY facilitates specification reuse, reduces cognitive effort, and helps prevent conformance issues early in the development process.

Furthermore, we report results from an empirical study that compares the use of DSL-REMEDY with a general-purpose DSL. The results show that REMEDY improves modeling productivity by reducing error rates, task completion time, and manual effort during PA specification. A complementary case study based on a robotic system illustrates the applicability of the method to realistic adaptive scenarios involving hierarchical feedback loops. In summary, REMEDY offers a structured and extensible solution for ACC in MAPE-K-based SASs, and provides insights relevant to broader adaptive software architectures.

This work extends our previous contribution~\cite{SanMartin2021} by evolving from a domain-specific language prototype into a comprehensive methodology for complete Architectural Conformance Checking. Specifically, the main contributions of this article are:

\begin{itemize}
    \item A revised and modular DSL metamodel that encodes MAPE-K abstractions and enforces domain-specific constraints;
    \item A model recovery tool that maps implementation-level artifacts to MAPE-K elements;
    \item A conformance checking engine that detects and visualizes structural mismatches between planned and implemented architectures;
    \item A realistic case study involving a robotic system with hierarchical feedback loops;
   \item A graphical visualization component that presents the Planned Architecture (PA), the Current Architecture (CA), and the detected architectural drifts, enabling clearer interpretation and decision-making.

\end{itemize}

Together, these contributions form a coherent method that addresses both specification and verification challenges in the architectural conformance of self-adaptive systems.

To facilitate traceability, the correspondence between the research questions and the article's sections is outlined below:

\begin{itemize}
    \item \textbf{RQ1} is addressed in \textbf{Section~\ref{support}}, particularly in Subsection~\ref{support:specification}, which introduces DSL-REMEDY and explains how it supports the precise definition of planned architectures using MAPE-K abstractions and embedded domain constraints.

    \item \textbf{RQ2} is answered in \textbf{Section~\ref{support}}, specifically in Subsection~\ref{support:conformance}, where we describe the conformance checking engine that compares the Planned and Current Architectures. This includes the use of OCL constraints and KDM representations to automatically detect architectural drifts.

    \item \textbf{RQ3} is addressed in \textbf{Section~\ref{evaluation}}, which presents a controlled experiment comparing DSL-REMEDY with a general-purpose DSL. The results demonstrate that DSL-REMEDY reduces modeling effort and improves precision.
\end{itemize}


Section~\ref{background} provides background on self-adaptive systems and architectural conformance checking. Section~\ref{case} introduces the case study that serves as a running example throughout the paper. Section~\ref{support} presents the REMEDY approach in detail, including its specification DSL, architecture recovery tool, and conformance checking engine. Section~\ref{evaluation} evaluates DSL-REMEDY through a controlled experiment. Section~\ref{related} reviews related work, and Section~\ref{conclusion} concludes the paper.

\section{Background} \label{background}

\subsection{Self-Adaptive Systems} \label{ass}

SASs are systems with the capability to adapt their behavior at runtime to change their execution conditions and user requirements \cite{Kephart2003}. Nowadays, they are highly demanded in pervasive, mobile, and embedded computing environments because their execution context requires a high degree of unpredictability and dynamism \cite{Weyns2023}.

SAS can be classified according to their adaptation goals. Therefore, self-configuring systems are those that  possess the capability to automatically configure themselves according to their operational requirements. Self-optimizing ones continually seek ways to improve their performance or reduce operational costs, ensuring efficiency and resource utilization. Self-healing systems are adept at identifying, diagnosing, and rectifying issues that arise from bugs or failures within the system, maintaining its operational integrity. Self-protecting systems can defend themselves against malicious attacks or mitigate the impact of cascading failures, ensuring the security and reliability of the system.

\begin{figure}[ht]
\centering
    \includegraphics[width=0.5\linewidth]{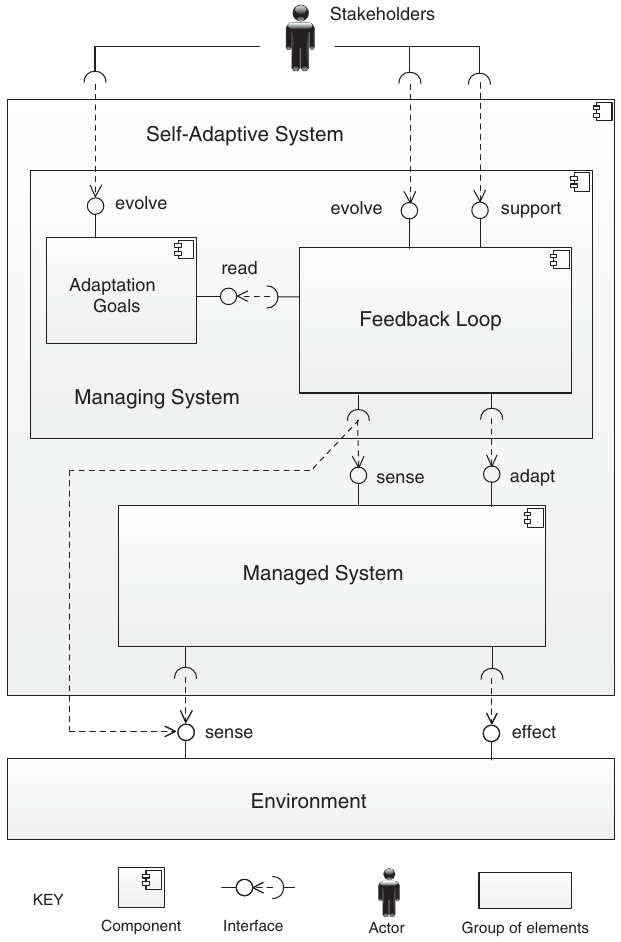}
    \caption{Conceptual model of a self-adaptive system from \citep{WeynsBook2021}}
    \label{fig:weyns}
\end{figure}

In 2005 IBM has published a reference model called MAPE-K for designing the feedback loop part of a SAS. MAPE-K prescribes some abstractions and the communication rules expected between these abstractions \cite{Ibm2005}. However as MAPE-K do not consider explicitly neither the system environment nor the adaptation goals, in 2020 \cite{WeynsBook2021} proposed a new broader conceptual model involving these concepts \cite{WeynsBook2021}. Figure \ref{fig:weyns} shows the Weyn's proposed conceptual model, highlighting the four core elements: the environment, the managed system, the feedback loop, and adaptation goals.

The environment denotes the external context where the SAS operates and its actions are assessed. It includes users, physical entities, and virtual components. The managed system embodies the application software responsible for executing the system's functions for its users, thus reflecting concerns related to the system's domain, or environment. To deliver its functionalities to users, the managed system interacts with and responds to the environment. To enable adaptations, the managed system must be equipped with sensors for monitoring and effectors (also known as actuators) for executing adaptation actions.

The managing system involves the adaptation goals and the feedback loop. The adaptation goals represent the objectives to be reached in the managed system by the actuation of the managing one, focusing on its quality attributes. Generally, four primary types of high-level adaptation goals are discernible.

\begin{figure}[ht]
\centering
      \begin{tikzpicture}[node distance=.8cm, legendtext/.style={text width=1.57cm, align=center}]
        \node (cal) at (0,-0.5) {\includegraphics[width=0.7\linewidth]{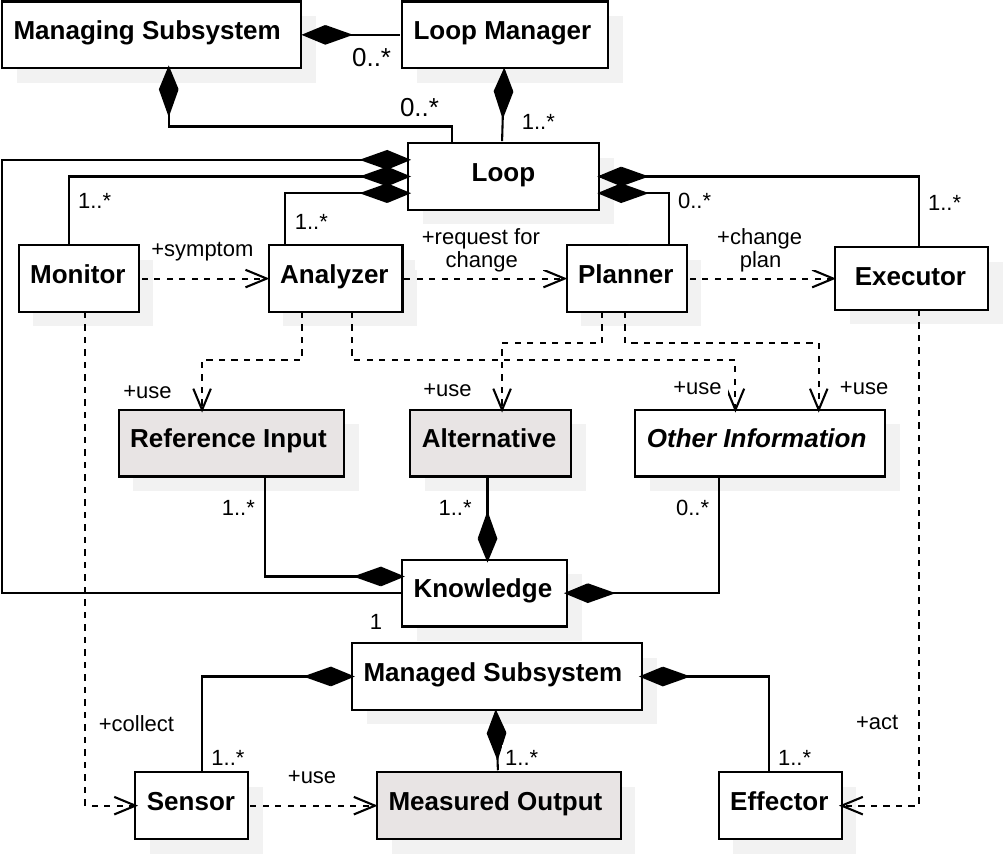}};
		\matrix [
				 draw=black,
				 fill=brown!20, 
				 column 1/.style={nodes={font=\tiny}},
				 column 2/.style={nodes={font=\tiny}},
				 column 3/.style={nodes={font=\tiny}}
				]
				at(0,4) {
		
			\node[fill=black,draw,diamond,inner sep=1.5pt,outer sep=0pt] (m) at (0,0){}; \draw (m) -- +(0,0.2); & 
			\node[draw,rectangle, fill=white, inner sep=3pt,outer sep=0pt] (o) {};& 
			\draw[-{Straight Barb[length=1mm,width=1mm, angle=60:5pt]},dashdotted] (0,-0.2)--(0,0.2);  \\
			\node[legendtext]{Composition}; & 
			\node[legendtext]{Abstraction}; &
			\node[legendtext]{Dependency}; \\
		};
   \end{tikzpicture}
   \caption{MAPE-K as an Abstract Planned Architecture}
   \label{fig:uml}

\end{figure}

Figure \ref{fig:uml} shows the MAPE-K metamodel enriched with some lower-level abstractions (in grey) presented by other works \cite{Villegas2011,Weyns2013, Abdennadher2017}. The elements Monitor, Analyzer, Planner, and Executor perform the four basic functions of any feedback loop and share a common Knowledge \cite{Muller2014, WeynsBook2021}. The Monitor abstraction acquires data from the managed subsystem and environment. The Analyzer abstraction determine whether there is a need for adaptation of the managed element. If adaptation is required, the Planner abstraction puts together a plan that consists of one or more adaptation actions. The adaptation plan is then executed by the Execute abstraction that adapts the managed element as needed \cite{WeynsBook2021}.

As previously said, besides the canonical abstractions, there are also some lower level abstractions that are not presented in the original representation of MAPE-K \cite{Villegas2011,Weyns2013}. These abstractions are in a lower-level of abstraction and they are essential when architects are working with detailed design of the system \cite{Villegas2011}. The grey boxes indicate abstractions that are not present in the standard MAPE-K. The \emph{Alternative} represents a set of available options that an SAS uses for changing the system behavior \cite{Abdennadher2017}. The \emph{Reference Inputs} consist of the concrete and specific set of values used to specify the state to be achieved and maintained in the managed system~\cite{Villegas2011}. The \emph{Measured Outputs} consist of the values that are measured in the managed system. Naturally, these measurements must be compared to the Reference Inputs to evaluate whether the desired state has been achieved~\cite{Villegas2011}.

\subsection{Architecture Conformance Checking}

Architectural Conformance Checking (ACC) is a fundamental activity in software quality assurance, aimed at detecting whether the implementation of a system deviates from the rules, constraints, and relations established in the Planned Architecture (PA)\cite{Knodel2007}. Such deviations, often referred to as architectural drifts, typically arise due to the gradual and often unconscious erosion of architectural integrity during system evolution\cite{Maffort2013,Mens2008}. A drift occurs whenever a reference model, or PA, prescribes certain rules, but the current implementation diverges from them, either unintentionally or as a consequence of pragmatic trade-offs. While architectural drifts do not always cause immediate system failures, they can significantly complicate maintenance, evolution, and overall system comprehensibility~\cite{Perry1992,Bass2015}. In severe cases, this phenomenon may escalate into architectural erosion or architectural decay, threatening long-term system sustainability~\cite{Kazman1998}.

A crucial step in ACC approaches is the explicit specification of the PA, which formalizes the architectural elements and the constraints governing their interactions. Two widely used techniques for PA specification are: \textit{(i)} reflexion models~\cite{Murphy1995}, and \textit{(ii)} constraint languages~\cite{Terra2009}. Reflexion models rely on the expertise of software engineers to define a high-level architectural abstraction, which is then mapped to a source model automatically extracted from the code. This facilitates the identification of mismatches between intended and actual dependencies. By contrast, constraint languages enable engineers to declaratively encode architectural rules using a generic vocabulary—such as components, modules, or subsystems—providing a more formal and analyzable representation of architectural design decisions~\cite{Rosik2011}. The REMEDY approach builds upon the latter technique, leveraging a domain-specific language (DSL) to precisely capture the core abstractions and rules of the PA, thereby enabling fine-grained control over dependency enforcement and violation detection~\cite{Maffort2013}.

\subsection{MAPE-K as a meta-reference model}

The MAPE-K shown in Figure \ref{fig:uml} represents the underlying metamodel we have adopted in REMEDY approach. So, it is an abstract Planned Architecture that must be instantiated for a specific system \citep{WeynsBook2021, Villegas2011, Abdennadher2017}. The names of abstractions and the relationships between them are exactly the same we have adopted in the DSL-REMEDY for specifying concrete planned architectures. 

The relationships that appear in the figure represent the rules that must be obeyed by a concrete planned architecture. They are the expected communication rules that must exist between the abstractions while the absence of them are the "denied/undesirable ones". For example, as there is no relation between Monitor and Planner, these elements are not expected to communicate. 

As this is a metamodel, different combinations of these abstractions are possible, providing flexibility for architects to decide the best combination. For example, it is perfectly possible to have a concrete PA with two or more Loops, with two or more Monitors or not including a Planner component.

Besides, this figure denotes two kinds of rules: \textit{i}) hierarchical rules and \textit{ii}) communication rules. Hierarchical rules denote composition of the abstractions. For example, Loop Managers can only be cretated/declared inside Managing Subsystems and they must aggregate Loops. Monitors, Analyzers, Planners, Executors and Knowledges can only exist inside Loops.  Sensors must be declared inside Managed Systems, and Alternatives must be declared inside Knowledge. 

There is an exception related to  Loops because they can be declared inside Managing Subsystems or inside Loops Managers. The latter occurs when two or more Loops (not necessarily deployed in the same location) need interaction to achieve the adaptation goal. Reference Inputs and Alternatives must be declared within Knowledge. Sensors, Measured Outputs and Effectors must be part of a Managed Subsystem, so these abstractions can be declared in this type of subsystems.

\section{Self-Adaptive Robot - A Case Study} \label{case}

\label{case-study}

This section introduces the PA of a self-adaptive robotic system utilized throughout the paper. The robotic system simulates adaptive behavior by adjusting its wheel parameters to maintain a desired distance from the wall, effectively mimicking a runtime feedback adaptation loop. While this represents a simplified control scenario, it serves as a suitable context for validating the proposed conformance checking methodology.

However, because of imperfections in the floor and wheels, the robot can not move in a straight line, i.e., it tends to get too close or too far from the wall, making a zigzag movement. Therefore, we design an adaptation module that constantly get the current distance from the robot to the wall and adjust the power and the steering angle of the wheels to put the robot back in its trajectory. 

Besides, our goal was also to make the robot to move with the best possible performance. That means the robot must be able to improve its correction parameters (power and steering angle) so that the zigzag movement becomes as narrow as possible.

\begin{figure}[ht]
\centering
    \includegraphics[width=0.8\linewidth]{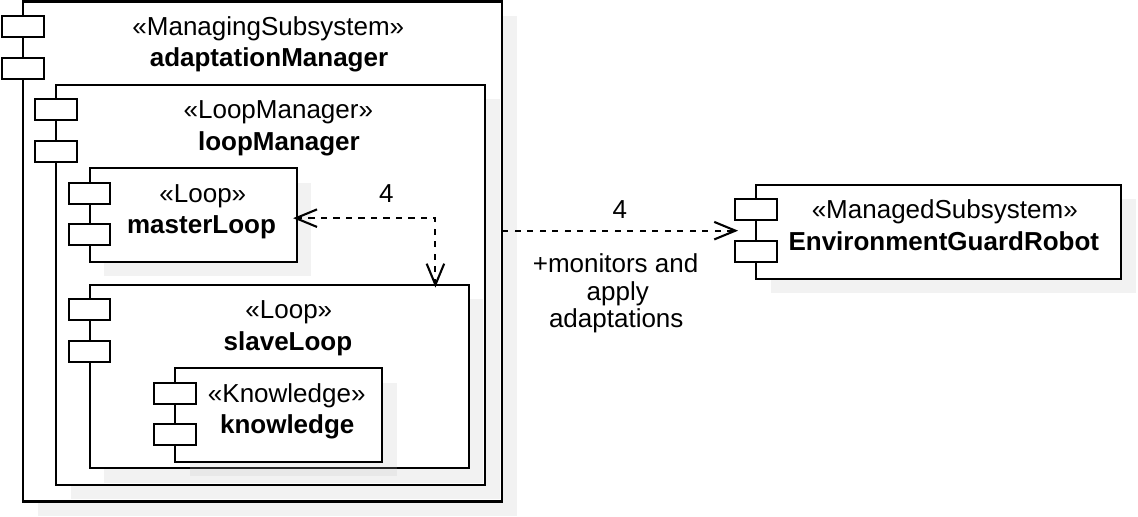}
    \caption{Concrete Planned Architecture - High Level View}
    \label{fig:high-level}
\end{figure}

\begin{figure}[ht]
    \centering
     \begin{tikzpicture}[node distance=.05cm, legendtext/.style={text width=1.5cm,align=center}]
        \node (cal) at (0,-1.4) {\includegraphics[width=0.7\linewidth]{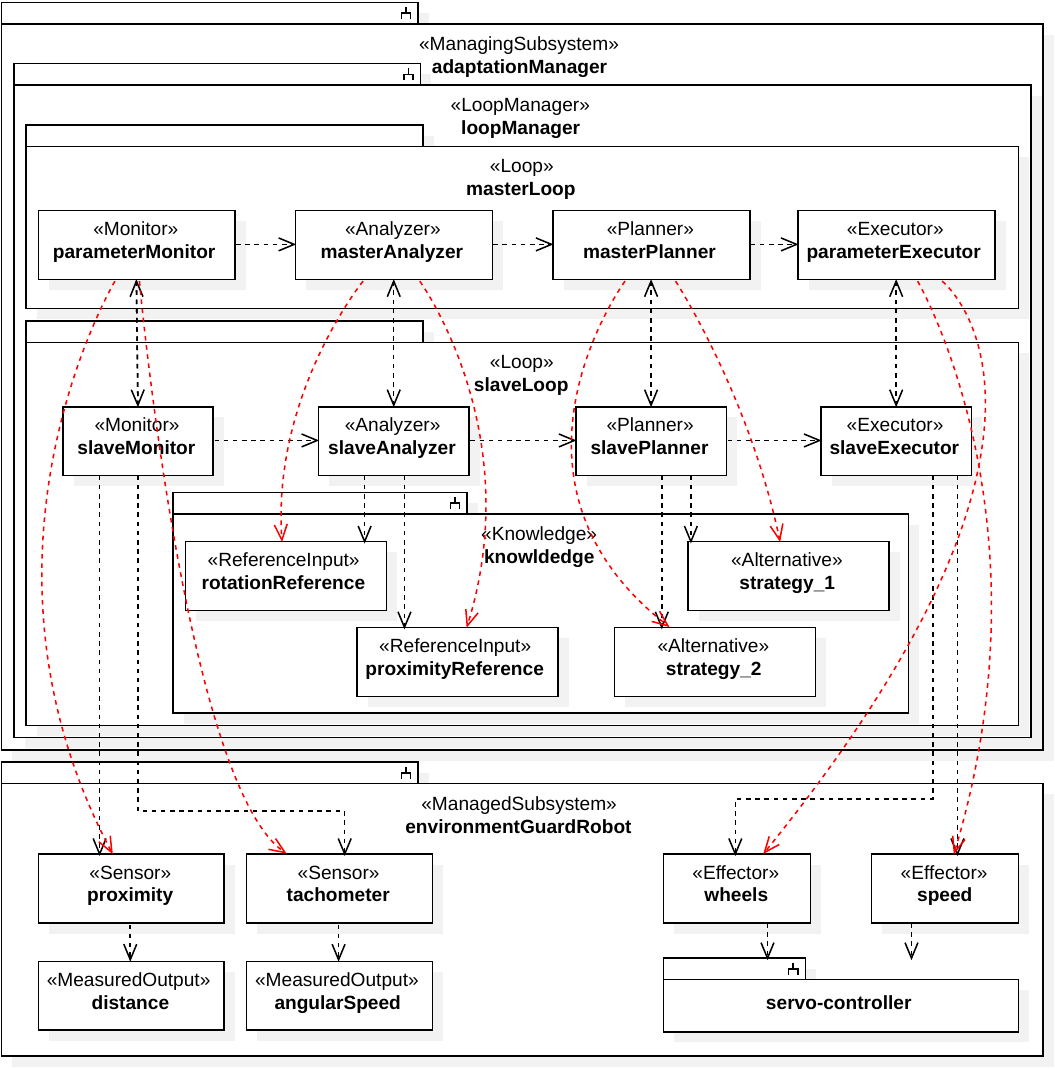}};
		\matrix [
				 draw=black,
				 fill=brown!20, 
				 column sep=0.3cm,
				 column 1/.style={nodes={font=\tiny}},
				 column 2/.style={nodes={font=\tiny}},
				 column 3/.style={nodes={font=\tiny}},
				 column 4/.style={nodes={font=\tiny}}
				]
				at(0,4) {
			[0ex]\draw[-{Straight Barb[length=1mm,width=1mm, angle=60:5pt]},dashdotted] (0,0)--(0.5,0);  & 
			[1ex]\node[fill=black,draw,diamond,inner sep=1.5pt,outer sep=0pt] (m) at (0,0){}; \draw (m) -- +(0,0.2); & 
			[0ex]\draw[red,-{Straight Barb[length=1mm,width=1mm, angle=60:5pt]},dashdotted] (0,0)--(0.5,0);  & 
		    [0ex]\node (r1) [draw,rectangle, fill= white, minimum width=0.4cm,minimum height=0.4cm]{};
		   	[0ex]\node[draw,rectangle, fill=white, below=1mm of r1.north, inner sep=3pt,outer sep=0pt] (o) {};\\
		   	
		   	[1ex]\node[legendtext]{must-use}; &
		   	[1ex]\node[legendtext]{composition}; &
		   	[1ex]\node[legendtext]{must-not-use}; &
		   	[1ex]\node[legendtext]{containment}; \\
		};
   \end{tikzpicture}
    \caption{Concrete Planned Architecture : A Low Level View }
    \label{fig:low-level}
\end{figure}

The PA for this robotic system is presented in two figures, a more high level view in Figure \ref{fig:high-level} and a more detailed one in Figure \ref{fig:low-level}. These graphical representations can be done in any UML-compatible tool and the intention is to formalize and also communicate the architectural decisions to the interested stakeholders. Notice that it is mandatory that each abstraction instance has a name, since it will be used in the next phases of the process. 

Figure \ref{fig:high-level} shows a high-level view of the concrete PA for this robotic system. The whole architecture consider the two basic subsystems; the Managed Subsystem representing the robot by itself and the Managing Subsystem representing the adaptation module. This module involves a Loop Manager and two Loops, one responsible for the most basic adaptation (Slave Loop) and another one responsible for adapting the slave one (Master Loop). The number of dependencies is indicated by a numeral displayed above the dependency arrows. Specifically, there are four dependencies between the Managing Subsystem and the Managed Subsystem, as well as four dependencies between the master and slave loops, in both directions.

This concrete PA was elaborated based on the elements available in the MAPE-K shown in Figure \ref{fig:uml}. Each package is an instance of a MAPE-K abstraction shown in Figure \ref{fig:uml}; the stereotypes denote the metamodel elements;

The focus os this view is on the main abstractions and in the hierarchical composition of the elements. The package \textit{adaptationManager} is an instance of the \textit{ManagingSubsystem}. The dependency between the \textit{adaptationManager} and the \textit{EnvironmentGuardRobot} means the first monitors and makes adaptations on the second. The package \textit{loopManager} is inside the package \textit{adaptationManager}. The ManagingSubsystem involves a LoopManager that, by its turn, involves two loops (Slave and Master) and a Knowledge component. The Slave Loop is responsible for regulating the robot's velocity and wheel adjustments. Meanwhile, the Master Loop oversees the first, analyzing and modifying adjustment parameters to enhance the robot's performance. 

Figure \ref{fig:low-level} shows the low-level view of the PA, focusing on the internals of the PA. Therefore, it is possible to see how the Loops were designed, the dependencies between them and also the relations with the Managed Subsystem. It is important to emphasize that the goal of this UML-based representation is to serve as a guide for software architects and also for communicating with the team. 

For the sake of conciseness, we have chosen not to provide an exhaustive elaboration of the Managed part, which typically exhibits a larger scope and greater complexity compared to the Managing part. As can be seen, the Managing aggregates just one Loop Manger \textit{loopManager}. The abstractions  \textit{ManagingSubSystem} and \textit{ManagedSubsystem} are \say{default} abstractions, as they must be presented in any specification.

In this configuration, the \textit{LoopManager} coordinates two feedback loops: \textit{masterLoop} and \textit{slaveLoop}. Each loop integrates the four canonical MAPE-K abstractions: Monitor (\textit{parameterMonitor, slaveMonitor}), Analyzer (\textit{masterAnalyzer, slaveAnalyzer}), Planner (\textit{masterPlanner, slavePlanner}), and Executor (\textit{parameterExecutor, slaveExecutor}). Beyond these, the \textit{slaveLoop} includes a \textit{Knowledge} element named \textit{knowledge}, which encapsulates two \textit{ReferenceInputs} (\textit{proximityReference} and \textit{rotationReference}) as well as two \textit{Alternatives} (\textit{strategy\_1} and \textit{strategy\_2}). These alternatives define distinct options that the system may exploit to adjust its behavior at runtime.

The \textit{ManagedSubsystem} complements this structure with environmental interaction mechanisms. It contains two \textit{Sensors} (\textit{proximity} and \textit{tachometer}), two \textit{Effectors} (\textit{wheels} and \textit{speed}), two \textit{MeasuredOutputs} (\textit{distance} and \textit{angularSpeed}), and a generic component termed \textit{servo-controller}, which is not explicitly represented in the DSL specification. The system establishes two categories of relations: \textit{communication rules}, illustrated as arrows, and \textit{structural relations}, shown through hierarchical composition. The latter indicate containment, i.e., when one abstraction is nested within another. As illustrated in Figure~\ref{fig:low-level}, the specification enforces $26$ \textit{must-use} rules and $8$ \textit{must-not-use} rules; whenever a communication rule is absent, the associated interaction is implicitly disallowed.

\section{The REMEDY Approach } \label{support}

\begin{figure*}[h]
	\centering
	 \begin{tikzpicture}
		\node (cal) at (0,-0.85) {\includegraphics[width=1\linewidth]{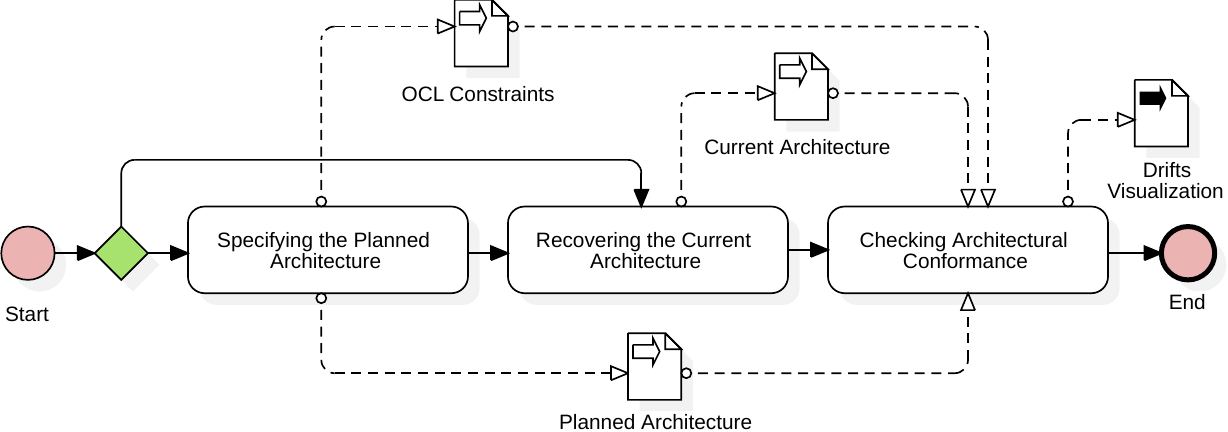}};
		\node at (-4.3,-0.9) {\textbf{A}};
		\node at (-0.9,-0.9) {\textbf{B}};
		\node at (2.5,-0.9) {\textbf{C}};

	\end{tikzpicture}
	
    \caption{The Three Steps of the REMEDY Approach}
    \label{fig:remedy}
\end{figure*}

REMEDY comprises three steps (A, B and C), as illustrated in Figure \ref{fig:remedy}. Each step is briefly explained below and detailed in the next subsections.

\begin{itemize}
    \item \textbf{Step A - \textit{Specifying the Planned Architecture}} is the process of creating a PA for a system using the DSL-REMEDY. This step can be conceptually divided in three sub-steps: i) specifying the Managing Elements; ii) specifying the Managed Elements and iii) specifying Communication Rules
                   
   \item \textbf{Step B - \textit{Recovering the Current Architecture}} is the process of identifying and the system’s as-is architecture based on its current implementation. This involves analyzing the source code elements, such as variables, methods, classes, and packages, and mapping them to the abstractions defined in the Planned Architecture (PA). The output of this step is an artifact that represents the current architecture (CA) of the system. This step is the moment when architects inform how the abstract elements are materialized in the source code. 

   \item \textbf{Step C - \textit{Checking Architectural Conformance}} is an automatic process responsible for identifying the drifts by comparing the PA with the CA. Internally, the PA and CA are represented as instances of the Knowledge Discovery Metamodel (KDM) \cite{Ulrich2010}.

\end{itemize}

\subsection{\textbf{Specifying the Planned Architecture}} \label{support:specification}

In this section we present the DSL-REMEDY and its application in the Robotic System we use as example. Figure \ref{fig:rr} shows the grammar that defines the structure of the DSL \cite{SanMartin2021}. At the core of this language are abstractions that operate on two levels: higher-level constructs for defining overarching architecture and lower-level components that specify detailed functionality.

\begin{figure}[ht]
\centering

\textbf{\scriptsize Higher-Level Abstractions}
\vspace{-0.3cm}
	\begin{grammar}
		\tiny
		\centering
		<ArchDefinition> ::=
		\begin{syntdiag}[\tiny\sdlengths]
			\lit{Architecture} \lit{ID} \lit{\{} \begin{rep}<ManagingSubsystem> \end{rep} \begin{rep}<ManagedSubsystem>   \end{rep}  \lit{\}} \lit{Rules}  \lit{\{}  \begin{rep} \begin{stack}   \\ <Rules> 	\end{stack} \end{rep}  \lit{\}}  \lit{;}
		\end{syntdiag} 
	\end{grammar}
	\vspace{-0.6cm}
		\begin{grammar}
			\tiny
			\centering
			<ManagingSubsystem> ::=
			\begin{syntdiag}[\tiny\sdlengths]
				\lit{Managing} \lit{ID} \lit{\{} \begin{rep} \begin{stack}<Loop>\\  <LoopManager>
				\end{stack} \end{rep}  \lit{\}}  \lit{;}
			\end{syntdiag} 
		\end{grammar}
	\vspace{-0.6cm}
		\begin{grammar}
		\tiny
		\centering
		<ManagedSubsystem> ::=
		\begin{syntdiag}[\tiny\sdlengths]
			\lit{Managed} \lit{ID} \lit{\{} \begin{rep} \begin{stack}<Sensor>\\  <Effector>\\ <MeasuredOutput>  \end{stack} \end{rep}  \lit{\}}  \lit{;}
		\end{syntdiag} 
	\end{grammar}
	\vspace{-0.6cm}
		\begin{grammar}
		\tiny
		\centering
		<Loop> ::=
		\begin{syntdiag}[\tiny\sdlengths]
			\lit{Loop} \lit{ID} \begin{stack} <DomainRule> \\ \end{stack} \lit{\{}  \begin{rep}
				\begin{stack} <Monitor>\\<Analyzer>\\<Planner>\\ <Executor>\\<Knowledge> \end{stack} 	\end{rep} \lit{\}}  \lit{;}
		\end{syntdiag} 
	\end{grammar}	
	\vspace{-0.6cm}
		\begin{grammar}
		\tiny
		\centering
		<LoopManager> ::=
		\begin{syntdiag}[\tiny\sdlengths]
			\lit{LoopManager} \lit{ID} \lit{\{} \begin{rep} <Loop> \end{rep}  \lit{\}}  \lit{;}
		\end{syntdiag} 
	\end{grammar}
		
\textbf{\scriptsize Lower-Level Abstractions}
\vspace{-0.3cm}
		\begin{grammar}
			\centering
		\tiny
		<Knowledge> ::=
	\begin{syntdiag}[\tiny\sdlengths]
		\lit{Knowledge} \lit{ID} \lit{\{} \begin{rep} \begin{stack} \\ <ReferenceInput> \\ <Alternative>					
			\end{stack}
		\end{rep} \lit{\}} \lit{;}
	\end{syntdiag} 
		\end{grammar}
\vspace{0.2cm}	

	\begin{minipage}[t]{0.4\linewidth}
		\begin{grammar}
			\tiny
			<Monitor> ::=
			\begin{syntdiag*}[\tiny\sdlengths]
				\lit{Monitor} \lit{ID}  \lit{;}
			\end{syntdiag*} 
			
			<Analyzer> ::=
			\begin{syntdiag*}[\tiny\sdlengths]
				\lit{Analyzer} \lit{ID}  \lit{;}
			\end{syntdiag*} 
			
			<Planner> ::=
			\begin{syntdiag*}[\tiny\sdlengths]
				\lit{Planner} \lit{ID}  \lit{;}
			\end{syntdiag*} 
			
			<Executor> ::=
			\begin{syntdiag*}[\tiny\sdlengths]
				\lit{Executor} \lit{ID}  \lit{;}
			\end{syntdiag*} 
			\vspace{-0.3cm}
		\end{grammar}
	\end{minipage}
	\hspace{0.3cm}
	\begin{minipage}[t]{0.5\linewidth}
		\centering
		\begin{grammar}
			\tiny
			<Sensor> ::=
			\begin{syntdiag*}[\tiny\sdlengths]
				\lit{Sensor} \lit{ID}  \lit{;}
			\end{syntdiag*} 
		
		<ReferenceInput> ::=
		\begin{syntdiag*}[\tiny\sdlengths]
			\lit{Effector} \lit{ID}  \lit{;}
		\end{syntdiag*} 
		
		<MeasuredOutput> ::=
		\begin{syntdiag*}[\tiny\sdlengths]
			\lit{MeasuredOutput} \lit{ID}  \lit{;}
		\end{syntdiag*} 
		
		<Alternative> ::=
		\begin{syntdiag*}[\tiny\sdlengths]
			\lit{Alternative} \lit{ID}  \lit{;}
		\end{syntdiag*} 
		\end{grammar}
	\end{minipage}
	
	\vspace{0.5cm}
\textbf{\scriptsize Rules}
	\bigbreak
	\begin{minipage}[b]{0.4\linewidth}
		\begin{grammar}
			\tiny
			<Rules> ::= 
			\begin{syntdiag*}[\tiny\sdlengths]
				\begin{rep}
					\begin{stack}  <RuleLoop> \\ <RuleMonitor> \\ <RuleAnalyzer>\\  <RulePlanner> \\ <RuleExecutor> \\ <RuleMeasuredOutput> \\ <RuleLoopManager> \\ <RuleKnowledge>  \end{stack}	\end{rep}
			\end{syntdiag*}
		\end{grammar}
	\end{minipage}
	\hspace{0.1cm}
	\begin{minipage}[t]{0.5\linewidth}
		\begin{grammar}
			\tiny
			\centering
			<RuleLoop> ::= 
			\begin{syntdiag*}[\tiny\sdlengths]
				\lit{loop} <Loop> \begin{stack} \lit{must-use} \\ \lit{must-not-use}	 \end{stack} \lit{loop} <Loop> \lit{;}
			\end{syntdiag*}
		\end{grammar}
		
		\begin{syntdiag*}[\tiny\sdlengths]
			\centering
			\tok{The other rules have a similar pattern so were omitted}
		\end{syntdiag*}
	
	\end{minipage}
\caption{ A simplified view of allowable structures of statements in REMEDY}
\label{fig:rr}
\end{figure}

At the higher level, the language begins with the definition of an architecture. This construct serves as the foundation, describing the system in terms of managing and managed subsystems, along with the rules that govern their interactions. The managing subsystem encapsulates entities responsible for coordinating and overseeing operations, often through constructs like loops and loop managers, which represent feedback mechanisms and their orchestrators. Meanwhile, the managed subsystem includes components that interact directly with the environment, such as sensors for data collection, effectors for action execution, and outputs that measure performance or results.

The notion of loops within the managing subsystem provides a formal way to define monitoring, analysis, planning, and execution processes. These loops are integral to the architecture, embodying the feedback-driven control mechanisms that adapt and respond to system dynamics. Supporting this, the concept of knowledge components ensures that the system maintains and leverages contextual and historical information during its operation.

At the lower level, the language refines these abstractions into granular constructs, such as the rules governing specific interactions or the detailed specification of sensors, effectors, and monitored outputs. This hierarchical approach allows for a seamless transition from abstract architectural design to detailed implementation, maintaining consistency and flexibility.

Finally, the grammar includes a rules section, which imposes constraints and guidelines on the interactions among the abstractions. These rules formalize relationships, dependencies, and restrictions, ensuring that the architecture adheres to the desired behavior.

\subsubsection{Using DSL-REMEDY for Specifying a Planned Architecture}

Listing \ref{lst:PAArch1} presents the specification of the Managing Subsystem depicted in Figure \ref{fig:low-level}. The composition of abstractions is written using a Java-like syntax. At this stage, the components of the Managing Subsystem must be declared. In our DSL, reserved words are highlighted in bold. Each abstraction can either be declared independently or composed of other abstractions, using the following notation:

\begin{center}
	[\begin{scriptsize}\ttfamily{SAS\_Abstraction}\end{scriptsize}] [\begin{scriptsize}\ttfamily{ID}\end{scriptsize}] [\begin{scriptsize}\ttfamily{\{..\}}\end{scriptsize}]; \quad | \quad [\begin{scriptsize}\ttfamily{SAS\_Abstraction}\end{scriptsize}] [\begin{scriptsize}\ttfamily{ID}\end{scriptsize}];
\end{center}

Here, [\begin{normalsize}\ttfamily{SAS\_Abstraction}\end{normalsize}] denotes a MAPE-K abstraction, and [\begin{normalsize}\ttfamily{ID}\end{normalsize}] is a globally unique identifier. When abstractions are nested inside others, they represent higher-level hierarchical compositions. The adoption of unique IDs facilitates referencing abstractions when specifying communication rules.

Line $1$ assigns a name to the PA. In this example, the system has a single Managing Subsystem (Line $2$) with one Loop Manager (Line $3$). A Loop Manager may coordinate one or more Loops; in this case, two Loops are defined (Lines $5$ and $12$). The \textit{masterLoop}, responsible for coordinating the \textit{slaveLoop}, contains four abstractions: \textit{parameterMonitor}, \textit{masterAnalyzer}, \textit{masterPlanner}, and \textit{parameterExecutor} (Lines $6$–$9$).

\lstset{style=sas} 
\begin{lstlisting}[caption= Managing subsystem of the PA,label=lst:PAArch1]
Architecture EnvironmentGuardRobot-PlannedArchitecture {
	Managing adaptationManager {
		LoopManager loopManager {
		
			Loop masterLoop withDomainRules{
				Monitor parameterMonitor;
				Analyzer masterAnalyzer;
				Planner masterPlanner;
				Executor parameterExecutor;
			}
			
			Loop slaveLoop withDomainRules{
				Monitor slaveMonitor;
				Analyzer slaveAnalyzer;
				Planner slavePlanner;
				Executor slaveExecutor;
				Knowledge knowledge {
					 ReferenceInput proximityReference;
					 ReferenceInput rotationReference;
					 Alternative strategy_1;
					 Alternative strategy_2;
...
}
\end{lstlisting}

Given the resource constraints of the robot, this specification promotes a decentralized organization of Loops, improving scalability in terms of communication and computation. For example, the \textit{masterLoop} could execute remotely on a server, while the \textit{slaveLoop} resides in the robot, thereby reducing communication overhead and distributing computational load \cite{Weyns2013}.

\lstset{style=sas} 
\begin{lstlisting}[caption= Managed subsystem of the PA,label=lst:PAArch2, basicstyle=\ttfamily\tiny,firstnumber=23]
Architecture EnvironmentGuardRobot-PlannedArchitecture {
..
	Managed environmentGuardRobot {
		 Sensor proximity;
		 Sensor tachometer;
		 Effector wheels;
		 Effector speed;
		 MeasuredOutput distance;
		 MeasuredOutput angularSpeed;
		 Component servo-controller;
		 
	}
}
..
\end{lstlisting}

The \textit{slaveLoop} is defined with its own \textit{Monitor}, \textit{Analyzer}, \textit{Planner}, \textit{Executor}, and an additional \textit{Knowledge} abstraction that contains reference inputs and alternative adaptation strategies, consistent with the DSL specification. Loops are declared using the \textbf{withDomainRules} keyword, which enables domain-specific constraints.

A complementary specification concerns the Managed Subsystem, which encapsulates the domain-level functionality. Interaction between Managing and Managed subsystems occurs through touchpoints, typically realized as sensors and effectors \cite{Ibm2005}. DSL-REMEDY allows these abstractions to be declared explicitly in the Managed Subsystem. Listing \ref{lst:PAArch2} illustrates this part of the PA specification.

In Line $25$, a Managed Subsystem is defined, which consists of two Sensors (Lines $26$–$27$), two Effectors (Lines $28$–$29$), two MeasuredOutputs (Lines $30$–$31$), and one generic component (Line $32$). These abstractions provide the required building blocks for modeling the Managed Subsystem.

Various approaches in the literature propose generic specification of systems using abstractions such as layers, components, or interfaces \cite{Terra2009, Jamshidi2013}. In our case, we extend the method proposed by Landi et al. \cite{Landi2017} and integrate it with our DSL. Since Landi’s framework also leverages KDM models to verify architectural conformance, it aligns well with the objectives of our approach.

The subsequent step is the definition of \textit{Communication Rules}. DSL-REMEDY adopts only two types of rules: \textit{must-use} and \textit{must-not-use}. The first indicates that abstraction \textit{A} is required to access abstraction \textit{B}, via method calls, object creation, interface implementation, etc. The second specifies that \textit{A} is prohibited from accessing \textit{B}.

\lstset{style=sas} 
\begin{lstlisting}[caption= Communication rules of the PA,label=lst:PARules, firstnumber=40]
Rules{
	loop masterLoop must-use loop slaveLoop;
	loop slaveLoop must-use loop masterLoop;
	monitor parameterMonitor must-use monitor slaveMonitor;
	monitor slaveMonitor must-use monitor parameterMonitor;
	monitor slaveMonitor must-use sensor proximity;
	monitor parameterMonitor must-not-use sensor proximity;
	monitor slaveMonitor must-use sensor tachometer;
	analyzer masterAnalyzer must-use analyzer slaveAnalyzer;
	analyzer slaveAnalyzer must-use analyzer masterAnalyzer;
	analyzer slaveAnalyzer must-use reference-input proximityReference;
	analyzer slaveAnalyzer must-use reference-input rotationReference;
	analyzer masterAnalyzer must-not-use reference-input proximityReference;
	analyzer masterAnalyzer must-not-use reference-input rotationReference;
	planner masterPlanner must-use planner slavePlanner;
	planner slavePlanner must-use planner masterPlanner;
	planner slavePlanner must-use alternative strategy_1;
	planner slavePlanner must-use alternative strategy_2;
	planner masterPlanner must-not-use alternative strategy_1;
	planner masterPlanner must-not-use alternative strategy_2;
	executor parameterExecutor must-use executor slaveExecutor;
	executor slaveExecutor must-use executor parameterExecutor;
	executor slaveExecutor must-use effector wheels;
	executor slaveExecutor must-use effector speed;
	executor parameterExecutor must-not-use effector wheels;
	executor parameterExecutor must-not-use effector speed;
	sensor tachometer must-use measured-output distance;
	sensor orientation must-use measured-output angularSpeed;
	effector wheels must-use Servo-Controller;
	effector speed must-use Servo-Controller;
}
\end{lstlisting}

Listing \ref{lst:PARules} provides the communication rules associated with Figure \ref{fig:low-level}. For instance, Lines $41$–$42$ define the mutual communication between \textit{masterLoop} and \textit{slaveLoop}. The DSL validates whether rules connecting both abstractions exist; otherwise, errors are reported during design time, and the PA cannot produce its output.

Conversely, if architects define a \textit{must-not-use} rule between Loops, all potential interactions between them are disregarded in the generated PA. Furthermore, multiple \texttt{LoopManagers} can be interconnected following the same principles.

Lines $43$–$47$ contain the rules concerning Monitors, $48$–$53$ refer to Analyzers, $54$–$59$ to Planners, $60$–$65$ to Executors, and $66$–$69$ to Sensors and Effectors.

Figure \ref{fig:rules} visualizes the set of allowable rules in DSL-REMEDY. Composable abstractions such as \texttt{LoopManager} and \texttt{Loop} are restricted to interact only with abstractions of the same type. For instance, \texttt{LoopManager} can only link to another \texttt{LoopManager}. Non-composable abstractions may also relate through \texttt{must-use} and \texttt{must-not-use} rules, but only within their own level. For example, a \textit{Monitor} may interact with other MAPE-K abstractions such as analyzers, planners, executors, or knowledge.

\begin{figure}[ht]
    \centering
    \tiny
    \begin{tikzpicture}[scale=0.6]
%
        \node (X) at (-19,0) [ellipse,draw,  minimum width=60pt] {Monitor};
        \node (W) at (-12.5,2) [ellipse,draw,  minimum width=60pt] {Analyzer};
        \node (U) at (-12.5,1) [ellipse,draw,  minimum width=60pt] {Planner};
        \node (T) at (-12.5,0) [ellipse,draw,  minimum width=60pt] {Executor};
        \node (S) at (-12.5,-1) [ellipse,draw,  minimum width=60pt] {Knowledge};
        \node (R) at (-12.5,-2) [ellipse,draw,  minimum width=60pt] {Sensor};

        \node (E) at (-8,-0.5) [ellipse,draw,  minimum width=60pt] {Analyzer};
        \node (D) at (-2,2) [ellipse,draw,  minimum width=60pt] {Monitor};
        \node (C) at (-2,1) [ellipse,draw,  minimum width=60pt] {Planner};
        \node (B) at (-2,0) [ellipse,draw,  minimum width=60pt] {Executor};
        \node (A) at (-2,-1) [ellipse,draw,  minimum width=60pt] {Knowledge};
        \node (F) at (-2,-2) [ellipse,draw,  minimum width=60pt] {Ref.Input};
        \node (G) at (-2,-3) [ellipse,draw,  minimum width=60pt] {Alternative};

        \node (I) at (-19,-5) [ellipse,draw,  minimum width=60pt] {Planner};
        \node (J) at (-12.5,-3) [ellipse,draw,  minimum width=60pt] {Monitor};
        \node (K) at (-12.5,-4) [ellipse,draw,  minimum width=60pt] {Analyzer};
        \node (L) at (-12.5,-5) [ellipse,draw,  minimum width=60pt] {Executor};
        \node (M) at (-12.5,-6) [ellipse,draw,  minimum width=60pt] {Knowledge};
        \node (N) at (-12.5,-7) [ellipse,draw,  minimum width=60pt] {Alternative};

        \node (O) at (-8,-6) [ellipse,draw,  minimum width=60pt] {Executor};
        \node (P) at (-2,-4) [ellipse,draw,  minimum width=60pt] {Monitor};
        \node (Q) at (-2,-5) [ellipse,draw,  minimum width=60pt] {Analyzer};
        \node (RR) at (-2,-6) [ellipse,draw,  minimum width=60pt] {Executor};
        \node (SS) at (-2,-7) [ellipse,draw,  minimum width=60pt] {Knowledge};
        \node (TT) at (-2,-8) [ellipse,draw,  minimum width=60pt] {Effector};

        \node (AA) at (-19,-9.5) [ellipse,draw,  minimum width=60pt] {Knowledge};
        \node (BB) at (-12.5,-8) [ellipse,draw,  minimum width=60pt] {Monitor};
        \node (CC) at (-12.5,-9) [ellipse,draw,  minimum width=60pt] {Analyzer};
        \node (DD) at (-12.5,-10) [ellipse,draw,  minimum width=60pt] {Planner};
        \node (EE) at (-12.5,-11) [ellipse,draw,  minimum width=60pt] {Executor};
        
        \node (FF) at (-8,-9) [ellipse,draw,  minimum width=60pt] {Sensor};
        \node (GG) at (-2,-9) [ellipse,draw,  minimum width=60pt] {Mea.Output};

        \node (Y) at (-8,-11.5) [rectangle,draw, minimum width=60pt] {Loop};
        \node (Z) at (-2,-11.5) [rectangle,draw, minimum width=60pt] {LoopManager};

        \draw[-latex] (Z) to [out=135,in=45,looseness=8] (Z);
        \draw[-latex] (Y) to [out=135,in=45,looseness=8] (Y);
     
        \path (X) edge [loop above] (X);
        \path (E) edge [loop above] (E);
        \path (I) edge [loop above] (I);
        \path (O) edge [loop above] (O);
        
        \draw[->] (X) -- ++(2,0) |- (W); 
        \draw[->] (X) -- ++(2,0) |- (U); 
        \draw[->] (X) -- ++(2,0) |- (T); 
        \draw[->] (X) -- ++(2,0) |- (S); 
        \draw[->] (X) -- ++(2,0) |- (R); 
    
       \draw[->] (E) -- ++(2,0) |- (D);
       \draw[->] (E) -- ++(2,0) |- (C);
       \draw[->] (E) -- ++(2,0) |- (B);
       \draw[->] (E) -- ++(2,0) |- (A);
       \draw[->] (E) -- ++(2,0) |- (F);
       \draw[->] (E) -- ++(2,0) |- (G);
       
       \draw[->] (I) -- ++(2,0) |- (J);
       \draw[->] (I) -- ++(2,0) |- (K);
       \draw[->] (I) -- ++(2,0) |- (L);
       \draw[->] (I) -- ++(2,0) |- (M);
       \draw[->] (I) -- ++(2,0) |- (N);
       
       \draw[->] (O) -- ++(2,0) |- (P);
       \draw[->] (O) -- ++(2,0) |- (Q);
       \draw[->] (O) -- ++(2,0) |- (RR);
       \draw[->] (O) -- ++(2,0) |- (SS);
       \draw[->] (O) -- ++(2,0) |- (TT);
       
       \draw[->] (AA) -- ++(2,0) |- (BB);
       \draw[->] (AA) -- ++(2,0) |- (CC);
       \draw[->] (AA) -- ++(2,0) |- (DD);
       \draw[->] (AA) -- ++(2,0) |- (EE);

        \path [->] (FF) edge node {} (GG);

        \path (X) edge [loop above] (X);
    
 \draw[dashed] (-21,-12.8) -- (1,-12.8);

        \node (composable) at (-17.5,-13.5)  [draw, rectangle] {\footnotesize Composable Abstraction};
       
       \node (noncomposable) at (-9,-13.5)  [draw, ellipse] {\footnotesize Non-Composable Abstraction};

        \node (relation) at (-1.5,-13.5)  [align=center, text width=3cm] {$\longrightarrow$: \footnotesize must-use/must-not-use};

\end{tikzpicture}

    \caption{The rules allowed by REMEDY. }
    \label{fig:rules}
\end{figure}

DSL-REMEDY also supports twenty predefined domain rules derived from the MAPE-K model, which can be activated or deactivated through the provided plugin. As an example, Planners must not access Monitors, and vice versa. Table \ref{tab:domainrules} lists all domain rules, where $\longrightarrow$ denotes \textit{must-use} and $\nlongrightarrow$ denotes \textit{must-not-use}. If a system violates these rules, we refer to it as domain drift. Note that in Listing \ref{lst:PARules}, domain rules were not explicitly declared, since they were already activated by default.

\begin{table}[ht]
\footnotesize
    \centering
     \caption{Domain rules of SASs}
    \label{tab:domainrules}
    \begin{tabularx}{\linewidth}{XXXXXX}
        \hline
        M &  A & P & E & K &\\
        \hline
        &$\nlongrightarrow$&$\nlongrightarrow$&$\nlongrightarrow$&$\nlongrightarrow$&M \\
       
        $\longrightarrow$&&$\nlongrightarrow$&$\nlongrightarrow$&$\nlongrightarrow$& A \\
      
        $\nlongrightarrow$&$\longrightarrow$&&$\nlongrightarrow$&$\nlongrightarrow$& P \\
        
        $\nlongrightarrow$&$\nlongrightarrow$&$\longrightarrow$&&$\nlongrightarrow$& E \\
        
        $\longrightarrow$&$\longrightarrow$&$\longrightarrow$&$\longrightarrow$&& K \\
         \hline  
    \end{tabularx}
   
\vspace{-4mm}
\end{table}

These rules are conceptual: they include both direct and indirect dependencies. A direct dependency occurs when an abstraction uses another directly in the CA. Indirect dependencies arise when an abstraction accesses another through an intermediary. In the latter case, REMEDY establishes the implicit dependency to ensure correct conformance checking.

\begin{lstlisting}[language=Xtend, caption={Checks itself abstraction dependency}, label={lst:checkmonitor}]
	if (dslRuleMonitor.monitor2 !== null)
	if (dslRuleMonitor.monitor == dslRuleMonitor.monitor2)
	error("Check a monitor does not have dependency with itself", SasDslPackage.eINSTANCE.DSLRuleMonitor_Monitor2, DUPLICATE_MONITOR_ACCESS)
\end{lstlisting}

\begin{lstlisting}[language=Xtend, caption={Checks the duplicity of a communication rule}, label={lst:checkduplicaterules}]
	for (entry:multiMapRuleMonitor2Monitor.asMap.entrySet) {
		val duplicates = entry.value
		if (duplicates.size > 1){
			for (d:duplicates)
			error("Duplicated rule",d, SasDslPackage.eINSTANCE.DSLRuleMonitor_Monitor2, DUPLICATE_RULES)
		}
	}
\end{lstlisting}

As noted earlier, the keyword \textit{withDomainRules} must be declared within Loops to enforce domain rules. Constraints are then applied to all abstractions affected by these rules. For example, if a Loop contains two monitors and the rule \textit{Monitor} $\nlongrightarrow$ \textit{Planner} is enabled, each monitor will be checked accordingly in the CA. While domain rules can be loaded automatically, architects may also introduce custom rules, even if they contradict existing ones. This highlights the importance of extending DSLs with validators capable of checking constraints beyond what parsing alone can ensure.

\begin{lstlisting}[language=Xtend, caption={Checks the violation of a domain rule}, label={lst:checkdomainrule}]
	var dslDomain = dslRuleMonitor.monitor.eContainer.eContents.filter(DSLDomainRule).toList
	if (!dslDomain.isEmpty) {
		val queryClass = new QueryClass(MainView.getDatabaseUrl())
		val rule = queryClass.ruleIsActive("Monitor","Planner");
		if (Boolean.valueOf(rule.get(1)))
		if (dslRuleMonitor.planner !== null && dslRuleMonitor.access.equals("must-use"))
		warning("The rule is violating the domain rule number  " + rule.get(0), SasDslPackage.eINSTANCE.DSLRuleMonitor_Planner)
	}
\end{lstlisting}

DSL-REMEDY implements three types of validators. The first ensures abstractions do not reference themselves (e.g., \textit{monitor parameterMonitor must-use monitor parameterMonitor} is invalid). Listing \ref{lst:checkmonitor} demonstrates this, where Line $1$ ensures the monitor is not null, Line $2$ checks for self-access, and Line $3$ raises an error.

Equivalent checks exist for the other abstractions and are not repeated here. The second validator prevents duplication of communication rules, implemented using a \textit{Hash} structure to track entries and raise errors if duplicates are detected (Listing \ref{lst:checkduplicaterules}). Finally, the third validator verifies compliance with domain rules, warning architects when a declared rule contradicts a deactivated domain constraint (Listing \ref{lst:checkdomainrule}).

\subsubsection{Artifacts Generated from the PA Step} \label{ocl-constraint}

In the first stage, the DSL produces two artifacts: (i) an OCL (Object Constraint Language) file that encodes the architectural rules declared in the Planned Architecture (PA), and (ii) a KDM (Knowledge Discovery Metamodel) instance that materializes the PA. The OCL artifact is later executed against the Current Architecture (CA) to check compliance, while the KDM model offers a language- and platform-independent representation suitable for analysis and visualization~\cite{KDM-ISO-2012}.

KDM is well-suited for our setting because it uniformly captures software elements at multiple abstraction levels, from source constructs to architectural components, allowing us to represent high-level PA abstractions alongside low-level CA entities within the same metamodel~\cite{KDM-ISO-2012}. 

We implement the generation pipeline with Xtend/Xtext templates derived from the DSL grammar~\cite{Xtext2013}. This choice aligns with the Eclipse/EMF-based ecosystem of REMEDY and streamlines code generation once the templates are defined. As a result, the PA is instantiated as a KDM model that can expose multiple viewpoints, ranging from high-level architectural structure down to source-level detail.

\begin{lstlisting}[caption= A PA Serialized as a Structure Package Instance,label=lst:kdmPA, language=xml]
	<kdm:Segment name="Planned Architecture">
	<model xsi:type="structure:StructureModel" name="ArchitecturalView_">
		<structureElement xsi:type="structure:Subsystem" name="adaptationManager" stereotype="/0/@extension.0/@stereotype.11">
				<structureElement xsi:type="structure:Component" name="loopManager" stereotype="/0/@extension.0/@stereotype.7">
				    [...]
						<structureElement xsi:type="structure:Component" name="parameterMonitor" stereotype="/0/@extension.0/@stereotype.0" outAggregated='//@model.1/@structureElement.0/@structureElement.0/@structureElement.0/@structureElement.0/@aggregated.0 [...] ' >
							<aggregated from='//@model.1/@structureElement.0/@structureElement.0/@structureElement.0/@structureElement.0' to='//@model.1/@structureElement.0/@structureElement.0/@structureElement.0/@structureElement.4' relation='//@model.0/@codeElement.0/@codeElement.1/@actionRelation.0 [...]' density='6'/> 
	  </structureElement>
		..
	</model>
	</kdm:Segment>	
\end{lstlisting}

Listing~\ref{lst:kdmPA} illustrates a fragment of the PA serialized as a KDM instance: 
Line~2 declares the architectural model; 
Line~3 defines the managing component (\textit{adaptationManager}); 
Line~4 specifies the loop component (\textit{loopManager}); and 
Line~6 declares the monitoring component (\textit{parameterMonitor}). 
Architectural relationships are expressed via the \textit{AggregatedRelationship} metaclass, which records links among elements. For example, Line~7 in Listing~\ref{lst:kdmPA} encodes the relationship between \textit{parameterMonitor}, \textit{slaveMonitor}, and \textit{masterAnalyzer}. 
Stereotypes are attached as lightweight extensions over KDM elements; they preserve the semantics needed for ACC without tying the model to a particular syntactic notation.

To support graphical exploration, we transform the KDM PA into a UML component diagram using a set of \textit{qvt-o} transformation rules. This model-to-model step underpins the visual inspection of the PA used later when reporting conformance results.

Listing~\ref{lst:ocl} shows an excerpt of the OCL file generated from Listings~\ref{lst:PAArch1}, \ref{lst:PAArch2}, and \ref{lst:PARules}, with domain rules activated for \textit{loop\_2}. Although the complete file exceeds 400 lines, its constraints are organized into four categories:
\begin{enumerate}
	\item \textbf{Existence rules}: ensure all declared abstractions are present.
	\item \textbf{Structural rules}: enforce containment and other structural constraints.
	\item \textbf{Communication rules}: validate permitted/forbidden accesses between elements.
	\item \textbf{Domain-specific rules}: apply additional constraints defined by architects.
\end{enumerate}

As a guide, the first rule (Lines~3–4) checks the existence and correct stereotype of \textit{parameterMonitor}; the second (Lines~6–8) verifies that \textit{parameterMonitor} is contained within \textit{masterLoop}; the third (Lines~10–11) forbids direct access from \textit{parameterMonitor} to \textit{proximity} (i.e., no \textit{AggregatedRelationship} between them); and the fourth (Lines~13–14) enforces the domain constraint $\text{Monitor} \nlongrightarrow \text{Planner}$, preventing an illegal connection between these abstractions.

\begin{lstlisting}[caption= Snippet of OCL constraints file,label=lst:ocl, language=OCL] 
package structure
-- Check SAS abstractions --
  context StructureModel
	inv exist_parameterMonitor: Component.allInstances()->exists(c| c.name='parameterMonitor' and c.stereotype->asSequence()->first().name = 'Monitor')
-- Check structural rules of SAS --
	context StructureModel
	inv composite_parameterMonitor: Component.allInstances()->select(c| c.name='parameterMonitor' and c.stereotype->asSequence()->first().name = 'Monitor')->
								  exists(d|d.oclContainer().oclAsType(Component).name='masterLoop' and d.oclContainer().oclAsType(Component).stereotype->asSequence()->first().name = 'Loop')
-- Check communication rules of SAS --
	context StructureModel
	inv not_access_parameterMonitor_proximity: not AggregatedRelationship.allInstances()->exists(c| c.from.name='parameterMonitor' and c.to.name='proximity')
-- Domain rules --
  context StructureModel
	inv domain_not_access_slaveMonitor_
	slavePlanner: not AggregatedRelationship.allInstances()->exists(c| c.from.name='slaveMonitor' and c.to.name='slavePlanner')
endpackage
\end{lstlisting}
\subsection{\textbf{Recovering the Current Architecture}}

In this paper we consider the CA as a model that involves: \textit{i)} the abstractions declared in the PA (without considering the communication rules) and \textit{ii)} the link between these abstractions and the source code elements that materialize them. In order to create such a model, software architects must map source code elements to the abstractions declared in the PA. The goal is to generate this model to that it can be compared to the PA. 

The mapping activity is facilitated when the system was designed with MAPE-K in mind. In this case, the names of MAPE-K abstractions are somehow reflected in the source code, what makes the identification easier. Naming conventions play a crucial role in simplifying the mapping process by establishing a direct correlation between architectural concepts and their implementation in the source code. For instance, a class named \texttt{MonitorHandler} or a package labeled \texttt{com.project.loop} intuitively links back to the MAPE-K abstractions \texttt{Monitor} and \texttt{Loop}. These conventions serve as initial clues that guide architects in identifying relevant elements.

In situations where naming conventions do not provide clear guidance, analyzing the semantic context of the code becomes essential. Architects must delve into the functionality and behavior of code elements to infer their mapping with architectural abstractions. For example, a method handling data aggregation might implicitly function as a \texttt{Monitor} or \texttt{Analyzer}, even if its name does not directly reflect this. This process often involves understanding the purpose of methods, classes, or modules by examining their interactions, dependencies, and roles within the larger system. Semantic analysis demands a deeper understanding of both the domain and the architectural principles, making it a critical yet challenging part of the mapping process.
	
\subsubsection{Making the mappings}

\begin{figure}[ht]
	
  \centering
\subfloat[Mappings for the Managed Subsystem]{
	\fbox{\includegraphics[width=1\textwidth]{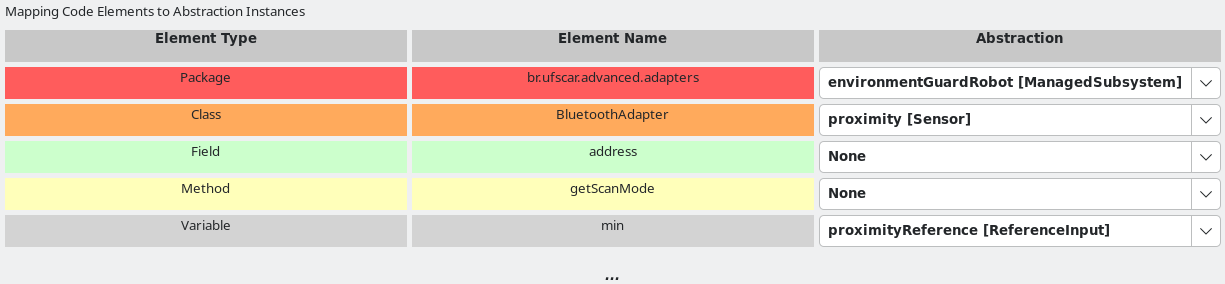}}
	\label{fig:annotation21}
}\quad
\subfloat[Mappings for the Managing Subsystem]{
	\fbox{\includegraphics[width=1\textwidth]{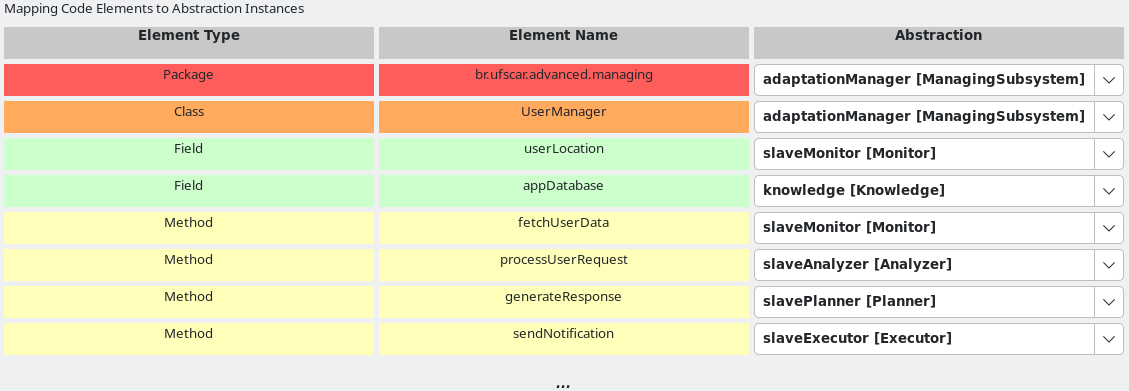}}
	\label{fig:annotation22}
}
 \caption{Computational support for mapping abstractions} 
\label{fig:mappingsupport}	
\end{figure}

Figure \ref{fig:mappingsupport} provides a schematic representation of the user interface of our Eclipse plugin for creating mappings. We have opted not to include a direct screenshot of the plugin due to its insufficient resolution, which would not be suitable for the reader.

The interface provides a tabular view with three columns: Element Type, Element Name, and the Abstraction to be assigned. The first two columns show source code elements of the class that is open in that moment in the Eclipse workbench. It is shown the element and its type, i.e., if it is a package, or class or field, etc. In the interface (a), the class \textit{BluetoohAdapter} is the class that is currently open in the Eclipse and the fields and methods of it are shown in the lines below. 

In the third column the architect can choose an abstraction instance to make the binding/assignment. The content of the dropdown field is shown in the format: \textit{name of the abstraction instance [type of the abstraction]}. 

Let us take as example the robotic system presented in Section \ref{case-study} and its Planned Architecture represented by the Listings \ref{lst:PAArch1}, \ref{lst:PAArch2} and \ref{lst:PARules}. For example, the variable \textit{min} was mapped to the \textit{proximityReference}, which is an instance of the \textit{ReferenceInput} abstraction. Notice that this abstraction instance was declared in the line 19 of the PA shown in Listing \ref{lst:PAArch1}.

Figure 8 shows two interfaces. The (a) presents some mappings for the Managed Subsystem while the (b) for the Managing one. Regarding the (a), notice that the package adapters was mapped to environmentGuardRobot, which is an instance of the Managed abstraction. 

The interface (b) shows mappings of the Managing Subsystem, which is the adaptive part, responsible for handling decision-making and coordination tasks within the robotic system. In this subsystem, elements such as the package \textit{managing} and class \textit{UserManager} are mapped to the \textit{adaptationManager} abstraction [Managing], highlighting their roles in processing data, generating responses, and managing overall system behavior.

As the plug-in works over the class that is currently open in Eclipse, the mapping process consists of navigating through the system opening the classes that must be mapped. Notice that, although a specific class must be open, it is also perfectly possible to map the package in which it is in. The types of code elements are highlighted with different colors for a better understanding. Also, the variables of a method (gray color) are represented just below of the method where they belong. For instance, the  \textit{min} variable belong to the method \textit{getScanMode}.

In our example the package \texttt{adapters} was mapped to the \texttt{environmentGuardRobot} abstraction; the class \texttt{BlueToothAdapter} to the \texttt{proximity} and the variable \texttt{min} to \texttt{proximityReference}. The abstraction names shown in the comboboxes are taken from the PA. 

\subsubsection{Technical Overview of Mappings}

The implementation of our plugin was done considering the KDM (Knowledge Discovery Metamodel) as the underlying metamodel for representing as the PA as the CA. 

Algorithm \ref{alg:CACreation} shows, in a high level of abstraction, the logic to create the Structure Model in the KDM of the CA. In Line $2$, the method creates the \textit{Structure Model} into the KDM with the SAS elements that were mapped previously. The architectural representation follows the same hierarchy as was implemented in source code. For instance, if a class and a field belonging to the same class were mapped with a \textit{Knowledge} and a \textit{ReferenceInput} respectively then our algorithm will create a component of type \textit{Knowledge} and inside it a \textit{ReferenceInput} component. To do so, we use the information stored in the embedded database and BaseX\footnote{https://basex.org/}, an XQuery engine, to manage XML based files.

\begin{algorithm}[ht]
\footnotesize
\caption{An overview of logical steps to create the structure package of a Current Architecture}
\label{alg:CACreation}
\begin{algorithmic}[1]
\REQUIRE \textit{KDM CA (kdmResource)}
\ENSURE \textit{KDM CA with an instance of the Structure Model}
\algrule
\FORALL {$packageMappings,classMappings,methodMappings,\newline fieldMappings,variableMappings$} 
\STATE $kdmResource \leftarrow createStructureElementFromTree()$
\STATE $kdmResource \leftarrow createImplementations()$
\STATE $kdmResource \leftarrow createRelationShips()$
\ENDFOR
\RETURN $kdmResource$
\end{algorithmic}
\end{algorithm}

The method $createImplementations()$ adds all source code elements that contribute to the implementation of an architectural element. For instance, it is possible to map two classes with the same abstraction and in that case both will implement the mapped abstraction. To do so, we use \textit{Query/View/Transformation operational (QVT-o)}, to perform the changes in the KDM CA.

In Line $4$ the method adds the relationships among SAS elements. REMEDY \cite{SanMartin2021} is capable of identifying $13$ types of relationships and as we explained before, they are created in the \textit{relation} attribute of the \textit{aggregated} metaclass. Thus, for every package, class, method, field and variable mapped to an SAS element the method identifies the relationships and adds them to a component or subsystem according to the case.

 \begin{lstlisting}[caption= A CA Serialized as a Structure Package Instance,label=lst:kdmCA, language=xml]
<model name="Current Architecture" xsi:type="structure:StructureModel">
  <structureElement name="adaptationManager" xsi:type="structure:Subsystem" implementation="//@model.0/@codeElement.0/@codeElement.0/@codeElement.0/@codeElement.0" outAggregated="" inAggregated="" stereotype="/0/@extension.0/@stereotype.12">
     <structureElement name="slaveMonitor" xsi:type="structure:Component" implementation="//@model.0/@codeElement.0/@codeElement.0/@codeElement.0/@codeElement.0/@codeElement.0/@codeElement.2" outAggregated="" inAggregated="" stereotype="/0/@extension.0/@stereotype.8"/>
     <structureElement name="slaveAnalyzer" xsi:type="structure:Component" implementation="//@model.0/@codeElement.0/@codeElement.0/@codeElement.0/@codeElement.0/@codeElement.0/@codeElement.1" outAggregated="" inAggregated="" stereotype="/0/@extension.0/@stereotype.9"/>
  ..
  </structureElement>
  ..
</model>
<model xsi:type="code:CodeModel" name="EnvironmentGuardRobot">
  <codeElement xsi:type="code:Package" name="br">
    <codeElement xsi:type="code:Package" name="ufscar">
      <codeElement xsi:type="code:Package" name="advanced">
        <codeElement xsi:type="code:Package" name="adapters">
           <codeElement xsi:type="code:Package" name="environmentGuardRobot">
              <codeElement xsi:type="code:ClassUnit" name="BluetoothAdapter" isAbstract="false">
 ..
 </model>
 \end{lstlisting}

Listing \ref{lst:kdmCA} presents a brief snippet of the CA generated by REMEDY. The snippet has been shortened for conciseness while preserving the overall structure. It consists of two models: the architecture model (line $1$), named Current Architecture, and the code model (line $9$). The architecture model is derived from mappings defined by the software engineer, whereas the code model is generated by the MoDisco plugin\footnote{\url{https://projects.eclipse.org/projects/modeling.modisco}}.

The architecture model represents three SAS abstractions: a managing subsystem (line $2$), a monitor (line $3$), and an analyzer (line $4$). Each abstraction includes attributes such as \textit{implementations}, \textit{outAggregated}, \textit{inAggregated}, and \textit{stereotype}. The \textit{implementations} attribute specifies the path in the KDM model where the corresponding code elements are located. The \textit{outAggregated} attribute describes the relationships between the architectural abstraction and others, including method calls and object creation. Conversely, the \textit{inAggregated} attribute captures incoming relationships from other abstractions. Finally, the \textit{stereotype} attribute classifies the abstraction type (e.g., monitor, analyzer).

As previously mentioned, the KDM model of the CA will be evaluated using multiple OCL constraints derived from step A to assess architectural conformance.

\subsection{\textbf{Checking the Conformance}} \label{support:conformance}

The architectural conformance checking process in REMEDY identifies deviations, referred to as drifts, between the Planned Architecture (PA) and the Current Architecture (CA) as implemented in the source code. This process generates a graphical representation of these drifts, visually highlighting inconsistencies such as missing architectural abstractions, violations of structural constraints, and deviations from prescribed communication rules.

From the user's perspective, this step is initiated when a software architect interacts with the Eclipse plugin by pressing a button in the UI. This action triggers the internal mechanisms of the workbench, which execute a series of algorithms to enforce OCL constraints on the CA. Algorithm \ref{alg:ACC} provides a high-level overview of this verification process. Additionally, a second algorithm compares the PA and CA KDM instances, computing the differences between the models. These differences are then visualized using third-party tools such as \textit{structurizr} and \textit{PlantUML}, providing a clear representation of the identified architectural drifts. Algorithm \ref{alg:emfcompare} provides a high-level overview of this process.

\subsubsection{Algorithm for Checking Architectural Constraints}

The algorithm for checking architectural constraints ensures that the current system implementation adheres to predefined architectural rules. It begins by initializing the OCL engine and loading the required resources, including the KDM instance representing the Current Architecture (CA) and a file containing the OCL constraints. Once these resources are prepared, the algorithm iterates through the constraint definitions, extracts their specifications, and organizes them into a mapping structure for efficient evaluation.

Next, the algorithm retrieves the structural model of the CA and systematically evaluates each constraint against it. The results of these evaluations are categorized based on the type of architectural rule they address, such as element existence, composite structures, access rules, or domain constraints. Finally, the categorized results are stored in a database, facilitating further analysis and visualization of potential architectural violations.

The algorithm follows these key steps:

\begin{enumerate}
	\item \textbf{Initialization and Resource Loading:} Line $1$ initializes the Pivot OCL engine, which evaluates constraints in MOF (Meta-Object Facility) models. Line $2$ creates an interface for interacting with the OCL engine. Line $3$ retrieves the KDM instance and loads it into memory. Line $4$ loads the OCL constraints into memory. Line $5$ initializes a HashMap to store the constraints, allowing efficient retrieval by key.
	\item \textbf{Constraint Extraction and Mapping:} Lines $6$–$14$ iterate over all objects in the constraint file, identifying those of type \texttt{Constraint}. For each constraint, the algorithm extracts its specification and stores it in the HashMap.
	\item \textbf{Structural Model Analysis and Constraint Evaluation:} Line $15$ retrieves the structural model of the KDM instance. Line $16$ iterates over all constraints stored in the HashMap. Line $17$ evaluates each constraint against the structure model of the CA, returning a boolean value, \texttt{false} if the constraint is violated and \texttt{true} otherwise.
	\item \textbf{Result Classification and Storage:} Lines $18$–$27$ categorize the evaluation results based on constraint type and store them in specific tables within the embedded database. This classification ensures structured storage and enables efficient retrieval for subsequent analysis.
\end{enumerate}

\begin{algorithm}[ht]
\footnotesize
\caption{Checking architectural constraints}
\label{alg:ACC}
\begin{algorithmic}[1]
\REQUIRE \textit{KDM CA, OCL constraintFile}
\algrule
\STATE $initOCL()$
\STATE $ocl \leftarrow OCL.newInstance$
\STATE $kdmResource \leftarrow resourceSet.getResource(CA)$
\STATE $oclResource \leftarrow ocl.parse(constraintFile)$
\STATE $constraintMap \leftarrow HashMap<String,ExpressionInOCL>$
\FORALL {$object \in oclResource$} 
\IF {$object$ \textbf{instanceof} $Constraint$}
\STATE $constraint\leftarrow object$ $\textbf{as}$ $Constraint$
\STATE $expression \leftarrow ocl.getSpecification(constraint)$
\IF {$expression$ \NOT $null$}
\STATE $constraintMap.put(expression.name, expression)$
\ENDIF
\ENDIF
\ENDFOR
\STATE $structureModel \leftarrow kdmResource.getStructure()$
\FORALL {$object \in constraintMap$} 
\STATE $check \leftarrow ocl.evaluate(structureModel, object)$
\IF {$object.key$ == $``exist"$}
\STATE $database.insertExistence(check,object)$
\ELSIF {$object.key$ == $``composite"$}
\STATE $database.insertComposite(check,object)$
\ELSIF {$object.key$ == $``access"$}
\STATE $database.insertAccess(check,object)$
\ELSIF {$object.key$ == $``domain"$}
\STATE $database.insertDomain(check,object)$
\ENDIF
\ENDFOR
\end{algorithmic}
\end{algorithm}

The checking algorithm reveals different classes of architectural drifts, each reflecting subtle discrepancies between the intended design and the current implementation. Missing abstractions emerge when the evaluation phase (lines $15–17$) cannot locate required elements defined in the constraint specifications, for instance, an analyzer or planner declared in the architecture but absent from the structural model of the current implementation. Structural violations are uncovered when the hierarchical relationships prescribed in the constraints are not respected, such as a planner being incorrectly embedded within a sensor abstraction. Communication drifts typically arise from the evaluation of access rules, where unauthorized interactions between monitors and effectors are detected, indicating a breach of the prescribed flow of information. Domain-specific inconsistencies, in turn, become visible when constraints tied to semantic properties, such as timing restrictions, execution order, or update frequency, are not satisfied, exposing deeper mismatches that may not manifest at the structural level alone.

These violations are not isolated findings but the result of systematic classification performed in the final stage of the algorithm (lines $18–27$). By categorizing each detected drift according to its rule type, existence, structure, access, or domain, the algorithm builds a coherent map of the inconsistencies, making it possible to distinguish recurrent patterns of erosion from isolated anomalies. The results are persistently stored in the embedded database of the workbench, ensuring that inconsistencies can be revisited, aggregated, and visualized over time. This persistent storage is particularly relevant for longitudinal analysis, where the evolution of architectural deviations can reveal whether violations are symptomatic of deeper design flaws or are merely transient effects of local modifications.

\subsubsection{Differences Between the PA and CA}

We compute differences between the PA and the CA using EMF Compare, which detects additions, deletions, modifications, and moves across EMF-based models. The comparison always begins at the root of both models to guarantee full coverage, since high-level mismatches often cascade into deeper inconsistencies in self-adaptive systems. Importantly, the results of this analysis are not treated as raw syntactic edits but are interpreted at the semantic level of SAS abstractions and architectural dependencies, ensuring that detected drifts are meaningful in architectural terms rather than superficial technical noise.

\begin{algorithm}[ht]
\footnotesize
\caption{Algorithm for comparing the PA and CA}
\label{alg:emfcompare}
\begin{algorithmic}[1]
\REQUIRE \textit{UML Package PA (resourceSet1), UML Package CA (resourceSet2), PlantUML PA File (file)}
\algrule
\STATE $initEMFCompare()$
\STATE $scope \leftarrow rangeOfComparison(resourceSet1,resourceSet2)$
\STATE $comparison \leftarrow compare(scope)$
\STATE $differences \leftarrow comparison.getDifferences()$
\FORALL {$difference \in differences$} 
\IF {$difference$ \textbf{is an} $ASElement$}
\IF {$difference$ \textit{is marked as DELETE}}
\STATE $listDeletedASElement \leftarrow difference$
\ENDIF
\ELSIF{$difference$ \textbf{is a} $Dependency$}
\IF {$difference$ \textit{is marked as DELETE} \textbf{and} $difference$ \textit{is marked as ADD}}
\STATE \textit{do nothing}
\ELSIF{$difference$ \textit{is marked as DELETE}}
\STATE $listDependency \leftarrow difference$
\ENDIF
\ENDIF
\ENDFOR
\FORALL {$ASElement \in listDeletedASElement$}
\STATE \textit{change the color of the AS element in file to gray}
\ENDFOR
\FORALL {$Dependency \in listDependency$}
\STATE \textit{change the color of the dependency in file to red}
\ENDFOR
\end{algorithmic}
\end{algorithm}

The process unfolds by first matching elements between the PA and the CA. This step is essential, because planners or monitors that have only been renamed or moved must not be misclassified as deletions and reintroductions. EMF Compare prioritizes stable identifiers when available, while falling back on names, roles, or structural positions when identifiers are missing. Once correspondences are established, the plugin derives the set of differences. At this stage, architectural drift manifests in two critical forms. On the one hand, when an abstraction mandated by the PA, such as a Monitor, Analyzer, Planner, or Effector, appears as deleted in the CA, this indicates erosion of the adaptive loop. On the other hand, the appearance of elements in the CA that have no PA counterpart may signal architectural creep. While some of these additions may be legitimate extensions, they frequently introduce inconsistencies that compromise the intended design and must be carefully scrutinized.

Dependencies require a more refined interpretation. A deletion immediately followed by an addition often corresponds to a refactoring, such as renaming an endpoint or relocating it across packages, rather than an actual loss of functionality. In such cases, the dependency is preserved semantically and reclassified accordingly. Only those dependencies that disappear without any replacement are recorded as genuine deletions, signaling the breakdown of a mandated interaction in the PA. This distinction prevents false positives while ensuring that truly critical communication links are not overlooked.

To improve comprehension, the inconsistencies are visualized on the PlantUML rendering of the PA. Deleted abstractions are displayed in gray, while missing dependencies are highlighted in red. These visual cues serve as more than decorative feedback: they accelerate diagnosis by making erosion patterns immediately visible. For example, the absence of a single Analyzer becomes perceptible as a structural gap in the adaptive loop, while the loss of a dependency is seen instantly as a break in communication flow. In this way, the visualization bridges the gap between syntactic differencing and architectural reasoning.

Several subtle cases deserve particular care. Rename and move operations frequently appear as delete–add pairs, but the interpretation reframes them as benign restructuring. Transient or generated elements in the CA can also create spurious differences, and these must be excluded by filtering known artifacts. The mismatch in granularity between PA (architectural level) and CA (implementation level) poses further challenges; here, mapping mechanisms and stereotyping rules are used to lift low-level changes to an architectural interpretation. Finally, the temporal persistence of differences provides additional diagnostic value: transient discrepancies often reflect ongoing refactoring, while sustained discrepancies reveal systematic erosion.

The differencing process provided by EMF Compare, enriched by semantic interpretation and visualization, yields a reliable account of architectural drifts. The deletion of SAS abstractions directly weakens the adaptive loop, while the disappearance of dependencies interrupts essential communication. By storing the detected drifts alongside the OCL-based violations discussed in Section~\ref{ocl-constraint}, we obtain a complementary perspective: one analysis captures violations of invariants that must hold, while the other uncovers concrete deviations over time. Together, they provide a deeper understanding of how self-adaptive architectures deviate from their planned specifications and why such deviations are critical for long-term architectural integrity.

\subsubsection{Visualizing Drifts - Technical View} \label{visualization}

The visualization of drifts constitutes a pivotal activity where software architects can visually the discrepancies between PA and CA. This visualization is made possible through model transformations performed on KDM instances, with the results utilized by a third-party tool named \textit{PlantUML}\footnote{https://plantuml.com/} for architectural visualization.

\begin{lstlisting}[language=QVT, caption= A QVT-o model transformation: from KDM to UML, label={lst:kdm2uml}]
mapping StructureModel :: Model2Model():Model when {self.name = modelName}{
	name :="UML Package Diagram";
	self.structureElement -> select(c | c.oclIsKindOf(Subsystem))[Subsystem] -> forEach(p){
		result.nestedPackage +=	p->map kdmSubsystem2UmlPackage();
	};
	
	self.structureElement -> select(c | c.oclIsKindOf(Component))[Component] -> forEach(p){
		result.nestedPackage +=	p->map kdmComponent2UmlPackage();
	}
}
mapping Subsystem:: kdmSubsystem2UmlPackage(): uml::Package {
	name := self.name;
	nestedPackage += self.structureElement -> select(c | c.oclIsKindOf(Component))[Component] ->  map kdmComponent2UmlPackage();
	self -> any(c| c.aggregated ->notEmpty()) -> aggregated -> forEach(t) {
	
		result.packagedElement += t-> map aggregated2dependency();		
	};
}
mapping Component:: kdmComponent2UmlPackage(): uml::Package {
	name := self.name;
	nestedPackage += self.structureElement -> select(c | c.oclIsKindOf(Component))[Component] ->  map kdmComponent2UmlPackage();
	self -> any(c| c.aggregated ->notEmpty()) -> aggregated -> forEach(t){ 
		result.packagedElement += t-> map aggregated2dependency();
	};
}
mapping  AggregatedRelationship::aggregated2dependency():uml::Dependency {
	client := self._from.late resolve(uml::Package);
	supplier := self.to.late resolve(uml::Package);
}
\end{lstlisting}

To streamline this process, the third-party tool Eclipse EMF Compare \footnote{https://www.eclipse.org/emf/compare/} is employed. This tool compares the PA and CA, both represented as UML Package diagrams, computes the differences between them, and generates a new model. This new model then undergoes the same transformation process for graphical visualization.

The transformation from KDM to UML considers just the structure model of KDM. The model is transformed into a UML Package Diagram. Listing \ref{lst:kdm2uml} shows the main parts of the QVT-o transformation file that takes as an input a KDM instance and returns as a result a UML instance.

Line $4$ and Line $8$ (mapping \textit{Model2Model()}) transform Subsystems and Components of the Structure Package of KDM into UML Packages. In our case the Structure Package is composed only of subsystems and components. Subsystems (the Managing and Managed abstractions) can only be composed of Components (others AS abstractions) and a Component can only be composed of Components.

Thus, in line $13$ (mapping \textit{kdmSubsystem2UmlPackage()}) the code transforms components that belong to a subsystem into UML Packages and it calls the mapping \textit{kdmSubsystem2UmlPackage()} to recursively transform the composition of components into UML packages. Similarly, in line $21$ it is done the same procedure if a Component is present in the model's high-level hierarchy. As in our representation, only components have relationships with other components it is necessary to implement a mapping that transforms the \textit{AggregatedRelationship} instances into UML Dependencies.

Line $27$ shows the mapping to transform the relationships of Structure Package of KDM into UML. In UML a dependency is composed by a \textit{client - supplier} relationship so the attribute \textit{from} is mapped to the attribute \textit{client} and attribute \textit{to} is mapped to the attribute \textit{supplier}. Notice that we use the special keyword \textit{late} to obtain a valid reference of the object to be transformed. A UML profile also is created to maintain a reference to the type of SAS abstraction in the UML Package.

Once the UML model is created, it is transformed again by using a specific API called \textit{structurizr \footnote{ https://structurizr.com/}} to convert the UML model into a DSL for being interpreted by PlantUML. The DSL generated is interpreted by PlantUML, a textual notation to construct UML diagrams.

\subsubsection{Visualizing Drifts - Functional View}

\begin{figure}[ht]
	\centering
	\begin{tikzpicture}
		\definecolor{newblue}{RGB}{52,152,219}
		\definecolor{newgray}{RGB}{192,192,192}
		
		\node (cal) at (0,-0.85) {\includegraphics[width=1\linewidth]{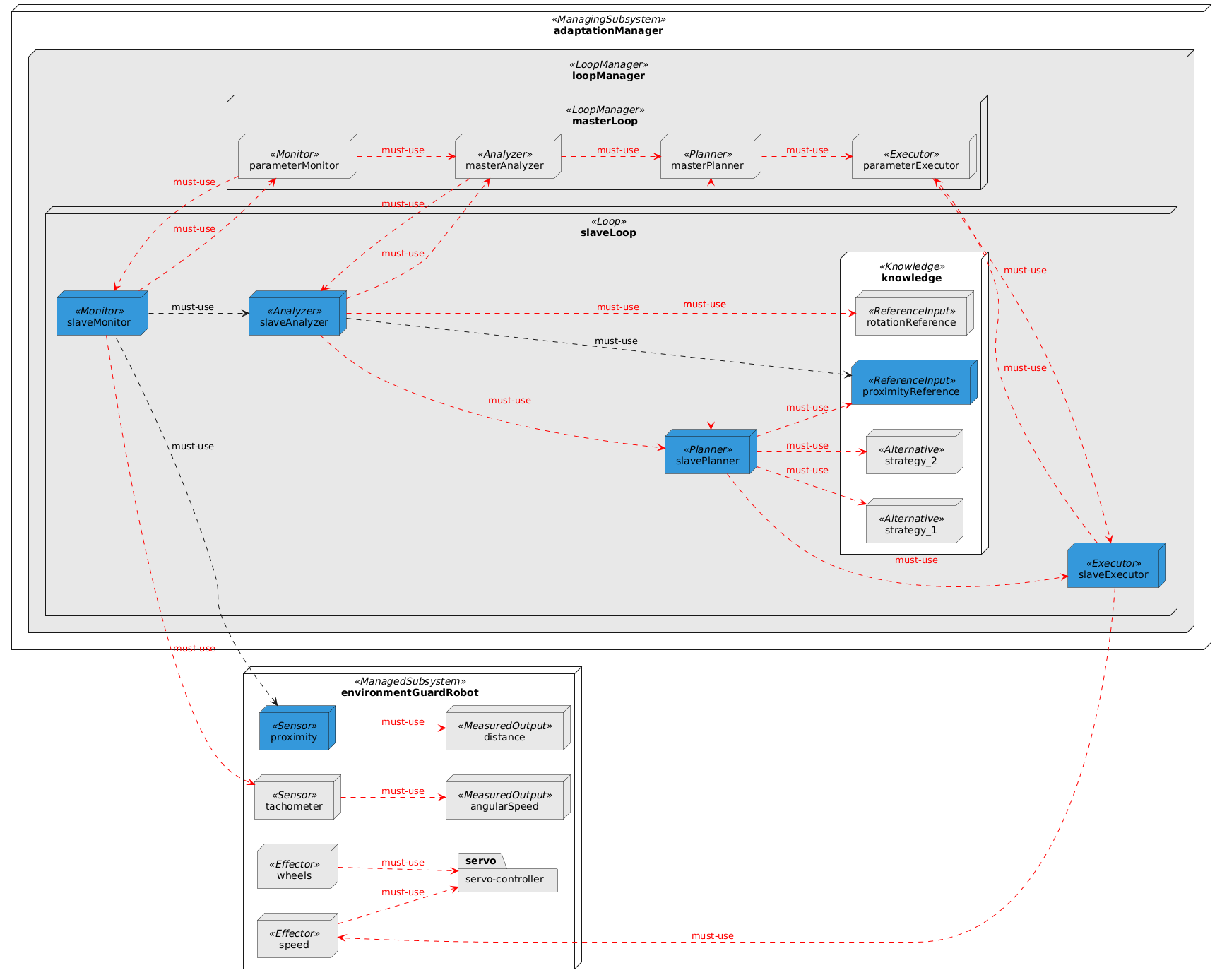}};
		\node[draw,fill=newblue,align=center] at (-0.15,-4) {\textbf{}};
		\node at (3,-4) {:\tiny non-composable abstractions present in the architecture};
		\node[draw,fill=newgray,align=center] at (-0.15,-4.3) {\textbf{}};
		\node at (2.4,-4.3) {:\tiny abstractions not included in the architecture};
		
		\node[draw,fill=white,align=center] at (-0.15,-4.6) {\textbf{}};
		\node at (2.75,-4.6) {:\tiny composable abstractions present in the architecture};
		
	 	\node at  (-0.15,-4.9) {\tikz \draw[red, thick, dash dot, -stealth] (0,0) -- (0.3,0);};
	 	\node at (0.4,-4.9) {:\scriptsize drift};
	 	\node at  (-0.15,-5.2) {\tikz \draw[black, thick, dash dot, -stealth] (0,0) -- (0.3,0);};
	 	\node at (0.95,-5.2) {:\scriptsize conformance};

	\end{tikzpicture}
	\caption{The visualization of drifts through structurizr and plantUML}
	\label{fig:tool9}
\end{figure}

Figure \ref{fig:tool9} illustrates an example of how REMEDY \cite{SanMartin2021} presents architectural conformance to software engineers. It is important to clarify that this is an example solely based on the mappings shown in Figure \ref{fig:mappingsupport} and corresponds to the robot case study. 

In this visualization, different symbols and arrows highlight where the implementation deviates from the intended MAPE-K architectural design. We explain the figure with respect to Existence Rules, Structural Rules, and Communication Rules, and discuss why detecting such drifts is crucial for maintainability and adaptability in SAS.
\noindent
\newline
A) Existence Rules: Missing Abstractions and Their Implications - One of the first things Figure \ref{fig:tool9} shows is the absence of certain architectural abstractions that were expected in the Planned Architecture but are missing in the implementation. In the visualization, REMEDY explicitly highlights these sixteen missing elements in gray color: \textit{loopManager}, \textit{masterLoop}, \textit{slaveLoop}, \textit{parameterMonitor}, \textit{masterAnalyzer}, \textit{masterPlanner}, \textit{parameterExecutor}, \textit{rotationReference}, \textit{strategy\_1}, \textit{strategy\_2}, \textit{tachometer}, \textit{wheels}, \textit{speed},  \textit{servo-controller},  \textit{angularSpeed} and  \textit{distance}.      

These correspond to MAPE-K components or domain concepts that the PA prescribed but the CA does not have. The implications of these missing abstractions are significant. Each abstraction in the PA was chosen to fulfill a role in the \textit{MAPE-K} loop or the environment interface, so omitting them in the implementation creates a gap between design and code. For instance, the absence of \textit{loopManager}, \textit{masterLoop}, and \textit{slaveLoop} removes the hierarchical structure that was intended to organize and coordinate adaptation processes. Without these elements, the feedback loops become implicit, potentially leading to ad-hoc adaptation logic scattered across the system instead of being modular and well-contained.

Similarly, missing adaptation components such as \textit{parameterMonitor}, \textit{masterAnalyzer}, \textit{masterPlanner}, and \textit{parameterExecutor} indicate that the \textit{master} adaptation process is not explicitly implemented, which weakens high-level decision-making. The absence of \textit{rotationReference} and strategies (\textit{strategy\_1}, \textit{strategy\_2}) suggests that predefined adaptation strategies and reference values are either being handled informally or not at all. Without these, the planner may lack structured guidance for adaptation, increasing the risk of inconsistent decision-making.

From a system interface perspective, missing environment and actuation components, such as \textit{tachometer}, \textit{wheels}, \textit{speed}, \textit{servo-controller}, and \textit{angularSpeed}, suggests that sensing and actuation mechanisms are either incomplete or have been implemented differently than planned. This omission implies that some actuator commands and sensor readings may be handled in an unstructured way, possibly bypassing the intended abstraction layers. Additionally, the absence of \textit{distance} as a measured output raises concerns about whether sensor data is being processed and stored correctly in the knowledge base or if it is instead being accessed directly by decision-making components. If the system is bypassing knowledge storage and retrieval, it could result in tight coupling between sensors and planners, making future modifications and expansions more difficult.

These \textit{existence-rule} violations indicate an erosion of the intended design: parts of the system’s conceptual architecture were not realized, which can confuse maintainers and lead to workarounds in the code. Over time, these gaps undermine the clarity of the adaptation logic and reduce the architectural integrity, the implementation no longer matches the original design, making it harder to verify, modify, or extend the system in a controlled manner.

\noindent
\newline
B) Structural Rules: Hierarchy and Containment Deviations - Structural rules define how components are organized, i.e., which components contain others and how the hierarchy of the MAPE-K loop is structured. The PA for this system envisioned a nested structure with an Adaptation Manager containing a LoopManager, which in turn contains two Loops (e.g., a master loop and a slave loop), each loop containing its Monitor, Analyzer, Planner, Executor and Knowledge abstractions.

Figure \ref{fig:tool9} reveals deviations in this hierarchy and containment structure when compared to the implementation. Because the \textit{loopManager}, \textit{masterLoop}, and \textit{slaveLoop} abstractions are missing in the code, the actual architecture does not mirror the intended nesting. In the PA, the \textit{loopManager} would act as a container and coordinator for two loops (master and slave), ensuring a clear separation between the two feedback loops and managing their interaction. Each \textit{Loop} abstraction would encapsulate its own \textit{MAPE-K} components (for example, a \textit{slaveLoop} containing \textit{slaveMonitor}, \textit{slaveAnalyzer}, \textit{slavePlanner}, \textit{slaveExecutor}, etc.). 

However, in the CA, with these container abstractions absent, the \textit{MAPE-K} components might all exist at one level or under an unintended package. For example, the \textit{slaveMonitor} and \textit{slaveAnalyzer} might exist in code but not inside a distinct \textit{slaveLoop} module, thus they are effectively floating in the architecture without the parent loop context that the design prescribed. The figure uses some visual cue such as a different color to show that those \textit{Monitors}, \textit{Analyzers}, etc., are present but their expected container (\textit{Loop}/\textit{LoopManager}) is not, indicating a structural drift.

Beyond the missing loop structures, the absence of key adaptation components such as \textit{parameterMonitor}, \textit{masterAnalyzer}, \textit{masterPlanner}, and \textit{parameterExecutor} further disrupts the structural organization. Without these elements, adaptation logic is likely fragmented rather than properly encapsulated within distinct monitoring, analysis, planning, and execution phases. Additionally, the absence of predefined strategies (\textit{strategy\_1} and \textit{strategy\_2}) suggests that planned decision-making processes may not have been implemented explicitly, potentially leading to a more reactive rather than structured adaptation mechanism.

The structural drift extends to missing environment and actuation components such as \textit{tachometer}, \textit{wheels}, \textit{speed}, \textit{servo-controller}, \textit{angularSpeed}, and \textit{distance}. These elements were intended to provide essential feedback for adaptation, particularly for controlling movement and adjusting based on environmental inputs. The absence of \textit{distance}, for instance, suggests that no explicit structure exists for handling proximity measurements in the knowledge base, which could result in components bypassing structured adaptation mechanisms and directly accessing sensor data.

This misalignment means the hierarchical relationships are broken: the intended containment (\say{part-of}) relationships in the architecture are violated. A direct consequence is reduced clarity in how the adaptive logic is organized. The separation between the \textit{master} and \textit{slave} adaptation concerns might be blurred in implementation, since without distinct loop entities, their components could be intertwined or managed collectively. Such structural drifts pose risks to maintainability because the modularity of the adaptation logic is compromised. If a developer expects to find all \textit{slaveLoop}-related logic grouped under a \textit{slaveLoop} module (per the architecture documentation) but instead finds it scattered, it takes more effort to locate and understand the behavior. It also becomes harder to reason about or replace one loop's functionality independently of the other. 

In SAS that rely on clear feedback loop boundaries, losing that structure could even impact adaptability – for instance, reconfiguring or scaling one loop (say, adding a new sensor to the \textit{slaveLoop}) might be harder if there is not an explicit structural separation. \textit{REMEDY} catching these structural rule violations allows engineers to see that the implementation's layering deviated from the plan. By visualizing that, say, \textit{Monitors} and \textit{Planners} are not under any \textit{Loop} container (because the container is missing), it prompts the realization that the code's organization needs refactoring or the architecture specification needs updating.

\noindent
\newline
C) Communication Rules: Expected vs. Unexpected Interactions - Communication rules govern which components are allowed to interact (call or data-flow relationships) in the \textit{MAPE-K} architecture. The PA defines allowed communications (for example, a \textit{Monitor} updates a \textit{Knowledge} base, an \textit{Analyzer} reads from that \textit{Knowledge}, a \textit{Planner} uses information from \textit{Knowledge} and issues commands to an \textit{Executor}, etc.) and forbidden ones (for example, a \textit{Planner} should not directly read a raw sensor, or a \textit{Monitor} should not directly command an \textit{Executor}, as these would bypass the layered responsibilities). In Figure \ref{fig:tool9}, \textit{REMEDY} visualizes the actual detected interactions in the code and flags any that violate the planned communication rules. Conforming interactions are likely shown with black arrows, whereas drift interactions are highlighted with red arrows.

A clear example of an unexpected interaction detected is the relation between \textit{slavePlanner} and \textit{proximityReference}. In the PA, there was no direct link intended between the \textit{slavePlanner} component and the \textit{proximityReference} element – any sensor data (such as \textit{distance} to an obstacle) needed by the planner was supposed to flow through the monitoring and knowledge abstractions. However, \textit{REMEDY} identified a \textit{slavePlanner} $\rightarrow$ \textit{proximityReference} coupling in the CA because it analyzes the relations that exist between the code elements that implement these architectural abstractions. In this case, the drift was detected due to \textbf{an object creation relationship} between the elements implementing \textit{slavePlanner} and \textit{proximityReference}. This indicates that the \textit{slavePlanner} in code is directly instantiating or maintaining a reference to \textit{proximityReference}, which was not part of the PA design. 

In Figure \ref{fig:tool9}, this shows up as a red arrow connecting the \textit{slavePlanner} element to the \textit{proximityReference} element, marking it as a drift (an unauthorized communication). This kind of drift suggests a developer shortcut or oversight: instead of retrieving the \textit{distance} via the \textit{Knowledge} (e.g., reading a \textit{distance} value that a \textit{Monitor} placed in the \textit{Knowledge} base), the planner is reaching out to \textit{proximityReference} directly through an object instantiation or reference, bypassing the intended data flow. Such an interaction breaks the \textit{MAPE-K} pattern by entangling the \textit{Planner} with low-level sensor details.

Beyond this specific case, \textit{REMEDY} would be checking other communication paths. For instance, if the architecture expects the \textit{slaveMonitor} to be the only one reading from the \textit{proximity} sensor, but the implementation had another component (like the \textit{parameterMonitor}, \textit{masterAnalyzer}, or an \textit{Executor}) also reading it, that too would be flagged. Or if an \textit{Executor} was invoking a \textit{Monitor}, etc., those would appear as deviations. The figure's red arrows (three of them are mentioned as identified drifts) call attention to each such unexpected coupling.

Other deviations may stem from the absence of expected communication paths due to missing abstractions such as \textit{loopManager}, \textit{masterLoop}, \textit{slaveLoop}, \textit{rotationReference}, \textit{strategy\_1}, \textit{strategy\_2}, \textit{tachometer}, \textit{wheels}, \textit{speed}, \textit{servo-controller}, \textit{angularSpeed}, and \textit{distance}. For example, if the \textit{slavePlanner} was supposed to get input from a \textit{Knowledge} element like \textit{distance}, an expected arrow from \textit{slavePlanner} to \textit{distance} might be absent because \textit{distance} itself is missing. The absence of \textit{loopManager}, \textit{masterLoop}, and \textit{slaveLoop} can also prevent the proper communication structure from being enforced, allowing unintended interactions to occur.

Each communication drift points to a potential design flaw or erosion – for example, a \textit{Planner} directly querying a sensor can make the system less modular (since changing that sensor or reusing the \textit{Planner} for a different context would be harder due to the direct dependency). Additionally, missing abstractions such as \textit{strategy\_1} and \textit{strategy\_2} suggest that predefined adaptation logic is either not implemented or is implicitly embedded within another component, reducing the separation of concerns that the PA originally intended.

Therefore the communication rules check in REMEDY distinguishes intended interactions from actual ones. Unexpected interactions like the \textit{slavePlanner–proximityReference} coupling are highlighted because they can lead to hidden dependencies and side effects. These violations may not crash the system, but they break the architectural guarantees about how information flows, which can reduce the clarity and analyzability of the adaptive behavior.

\section{Evaluation of DSL-REMEDY} \label{evaluation}

This section presents an empirical study to assess the effectiveness of REMEDY’s DSL in supporting the specification of planned architectures for MAPE-K-based SASs. The goal is to evaluate how well DSL-REMEDY \cite{SanMartin2021} supports productivity and precision when defining architectural models for self-adaptive systems compared to a general-purpose alternative.

The evaluation presented here differs from our earlier work~\cite{SanMartin2021} in three ways: (i) the DSL used in the current study has been redesigned to incorporate a broader metamodel with domain constraints; (ii) the conformance checking engine introduced here was not present in~\cite{SanMartin2021}; and (iii) although the participant pool remains the same (software engineering students), the tasks now include architectural violation detection rather than just modeling precision, thus assessing different capabilities.

\subsection{Scoping}

One of the claimed benefits of using domain-specific ACC approaches is the improvement of productivity, since some structural and communication rules from the domain are already known and therefore do not need to be specified explicitly by software architects \citep{Elizondo2017}. Nevertheless, to the best of our knowledge, no controlled experiments have been conducted that provide concrete evidence of productivity gains when software architects employ domain-specific ACC approaches, particularly in the research area of self-adaptive systems (SASs).  

Hence, the contribution of this section is to present a controlled experiment designed to compare two different approaches and tools for Architectural Conformance Checking. The first is DCL-KDM \citep{Landi2017}, a generic ACC approach, and the second is DSL-REMEDY, a domain-specific ACC approach tailored for SASs. The goal of our experiment is formally stated as follows:

\begin{table}[ht]
	\scriptsize
	\centering
	\begin{tabular}{|l|}
		\hline 
		\textbf{Analyze} the architecture specification of SASs with ACC tools $\dagger$ \\
		\textbf{for the purpose} of evaluating two different tools \\
		\textbf{with respect to their} productivity \\
		\textbf{from the point of view of} researchers \\
		\textbf{in the context of} final-year undergraduate students in computer engineering. \\
		\hline 
		$\dagger$ Refers specifically to the adaptive part of an SAS.\\
		\hline
	\end{tabular}
\end{table}

In this context, \textit{productivity} refers to three main factors: the cost in time required to complete the specification, the number of errors introduced during the process, and the overall work effort demanded of the participants. The experiment was conducted within the scope of a PhD research project, involving final-year students of the Computer Engineering Bachelor program at a well-known Chilean university. This population was chosen because, although they were still students, they had sufficient training in software architecture to perform the assigned tasks in a realistic and controlled setting.

\subsection{Setting}

The experiment was conducted over the course of one week at the end of the second semester of 2020. The participants were 24 final-year undergraduate students who had already completed courses in software engineering, domain-specific languages, and software architecture. This background provided them with a solid foundation in the fundamental concepts and principles relevant to designing and developing software systems, including adaptive systems. The experiment consisted of three activities, each requiring four hours of work per day:  

\noindent
1. \textbf{Training session.} This stage introduced both theoretical topics, such as self-adaptive systems, software architecture, and DSLs, and practical aspects focused on the use of REMEDY and DCL-KDM. During this session, students also completed a profiling questionnaire that gathered information about their programming skills, industry experience, and academic coursework related to the experiment. The results of this questionnaire were used to divide them into two balanced groups of 12 students each.  

\noindent
2. \textbf{Pilot experiment.} In this phase, students signed consent forms and then worked in groups to specify the architecture of two self-adaptive systems, one using REMEDY and the other using DCL-KDM. The pilot allowed us to assess whether the allocated time was sufficient, whether the instructions were clear, and whether the tools were being applied correctly.  

\noindent
3. \textbf{Main experiment.} Finally, the actual experiment was carried out, following the same format as the pilot but using different architectural specifications of self-adaptive systems.  

REMEDY is available at \url{https://tinyurl.com/y34jyeut}, and DCL-KDM can be accessed at \url{https://github.com/dedeLandi/dcl-kdm}.

\subsection{Planning}

\subsubsection{Experimental Design} 
Table \ref{tab:scheme} presents the design of the experiment. Students were divided into two balanced groups according to their profiles, ensuring that prior experience and background characteristics were evenly distributed. This balancing process reduces the risk of bias, since significant differences between groups could introduce confounding variables and compromise the validity of the results. Each group was assigned two different tasks involving the specification of self-adaptive systems (SASs).  

The assignments were presented as hypothetical case scenarios described using UML notation. Although hypothetical, both tasks were intentionally designed to be of equal complexity, drawing on well-known patterns of architectural styles for SASs \citep{Weyns2013}. Each specification required the definition of SAS abstractions and their associated communication rules, creating a consistent architectural framework for evaluation.  

\begin{table}[ht]
	\scriptsize
	\centering
	\caption{Experiment design}
	\label{tab:scheme}
	\begin{tabularx}{1\textwidth}{XXX}
		\hline  
		\textbf{Group} & \textbf{First Task}& \textbf{Second Task } \\
		\hline
		G1 & S-I (DCL-KDM) &   S-II (REMEDY)\\
		G2 & S-II (REMEDY) &   S-I (DCL-KDM)\\
		\hline  
	\end{tabularx}
	\vspace{-4mm}
\end{table}

In the first task, Group~1 specified S-I using DCL-KDM while Group~2 specified S-II using REMEDY. In the second task, Group~1 specified S-II with REMEDY and Group~2 specified S-I with DCL-KDM. The complete set of specification diagrams, datasets, and R scripts used for statistical analysis are publicly available at \url{https://doi.org/10.5281/zenodo.8175111}.  

This experimental design ensures fairness and replicability by subjecting both tools to identical conditions. Real-world systems vary considerably in complexity and constraints, which makes direct comparisons problematic. By employing hypothetical yet balanced specifications, the evaluation isolates the effectiveness and efficiency of the tools themselves. This controlled setup provides a clear understanding of their strengths and weaknesses, eliminates noise from domain-specific artifacts, and enhances transparency and reproducibility of the findings.

\subsubsection{Hypotheses Formulation} 
The research goal of the experiment is to compare DCL-KDM and REMEDY in terms of productivity, specifically focusing on the time required to complete an SAS specification, the number of errors introduced, and the perceived effort reported by participants. Accordingly, three null hypotheses ($H_0$) and three corresponding alternative hypotheses ($H_{\alpha}$) were formulated:

\begin{itemize}
	\item $H_{01}$: There is no difference in the time required to complete an SAS specification when using DCL-KDM or REMEDY.
	\begin{align}
		H_{01}:& \mu_{time_{DCL-KDM}} = \mu_{time_{REMEDY}}\\
		H_{\alpha1}:& \mu_{time_{DCL-KDM}} > \mu_{time_{REMEDY}} 
	\end{align}
	
	\item $H_{02}$: There is no difference in the number of errors made when specifying an SAS with DCL-KDM or REMEDY.
	\begin{align}
		H_{02}:& \mu_{errors_{DCL-KDM}} = \mu_{errors_{REMEDY}}\\
		H_{\alpha2}:& \mu_{errors_{DCL-KDM}} > \mu_{errors_{REMEDY}} 
	\end{align}
	
	\item $H_{03}$: There is no difference in the perceived effort when specifying an SAS with DCL-KDM or REMEDY.
	\begin{align}
		H_{03}:& \mu_{effort_{DCL-KDM}} = \mu_{effort_{REMEDY}}\\
		H_{\alpha3}:& \mu_{effort_{DCL-KDM}} > \mu_{effort_{REMEDY}} 
	\end{align}
\end{itemize}

\subsubsection{Independent and Dependent Variables} 
Each hypothesis requires the definition of independent and dependent variables, along with appropriate metrics for measurement.  

\textit{Independent variables} are those manipulated within the experiment. In this study, two were considered:  
- \textbf{Technique:} the tool used to complete the assignment, with two possible values: DCL-KDM and REMEDY.  
- \textbf{Specification:} the architectural problem to be solved (S-I or S-II). Both were intentionally designed with the same level of difficulty, so while this variable influences the task, it is treated as fixed rather than as a factor.  

\textit{Dependent variables} capture the outcomes of interest. For each hypothesis, the corresponding dependent variables were defined as follows:  
- \textbf{Time:} measured in minutes to complete the SAS specification.  
- \textbf{Errors:} the number of mistakes detected after the specification was completed.  
- \textbf{Effort:} the subjective perception of difficulty, measured on a Likert-type scale from $1$ (very easy to use) to $4$ (very difficult to use).  

\subsection{Analysis \& Discussion}

\subsubsection{Analysis}

A total of $24$ subjects produced valid data for paired comparisons across the three dependent variables: time, errors, and perceived effort. Table~\ref{tab:data1} summarizes the means and standard deviations for each variable under the factor Tool. While descriptive statistics already suggest notable differences between DCL-KDM and REMEDY, statistical testing was conducted to assess whether these differences were significant.  

\begin{table}[ht]
	\scriptsize
	\centering
	\caption{Mean and standard deviation of time, error, and effort}
	\label{tab:data1}
	\begin{tabularx}{\linewidth}{XXXXXXX}
		\hline  
		& \multicolumn{2}{c}{\textbf{Time}} & \multicolumn{2}{c}{\textbf{Error}} & \multicolumn{2}{c}{\textbf{Effort}} \\
		\hline
		\textbf{Tool}& \textbf{mean} & \textbf{sd }& \textbf{mean} & \textbf{sd }& \textbf{mean} & \textbf{sd }\\
		\hline
		DCL-KDM& 64.75& 18.43024 & 4.791& 3.106 & 2.83& 0.637 \\
		REMEDY & 45.75 & 16.67920 & 1.291 & 1.122& 1.70 & 0.624 \\
		\hline 
	\end{tabularx}
	\vspace{-4mm}
\end{table}

Because the design involved one factor with two levels in a within-subjects setting, paired-sample tests were applied. To avoid carryover effects, a full counterbalancing strategy was used (see Table~\ref{tab:scheme}). A t-test on orders yielded a p-value of $0.9174$, indicating that task order had no significant effect on performance. Normality of residuals was confirmed with Shapiro–Wilk tests for subjects and subjects$\times$tools ($p=0.1119$ and $p=0.06314$, respectively). A paired t-test on tools then showed a significant difference in time ($p=0.0004403$), demonstrating that the completion time differed systematically between DCL-KDM and REMEDY.  

For errors, the distribution under DCL-KDM conformed to a Poisson model ($p=6.89\mathrm{e}{-06}$), while the REMEDY data did not, motivating the use of non-parametric analysis. A Wilcoxon signed-rank test confirmed significant differences ($p=5.722\mathrm{e}{-6}$), with DCL-KDM producing consistently more errors.  

Effort, measured on a Likert scale from $1$ to $4$, was also analyzed with the Wilcoxon signed-rank test. The test yielded $p=4.768\mathrm{e}{-07}$, indicating a clear difference in perceived difficulty between the two tools. In all three cases, the null hypotheses were rejected at $\alpha = 0.05$.  

\subsubsection{Discussion}

The results demonstrate that REMEDY offers substantial advantages over DCL-KDM in terms of productivity and usability. On average, participants required approximately $30\%$ less time to complete specifications with REMEDY. This improvement is likely explained by the availability of predefined domain-specific rules aligned with the MAPE-K model, which simplified specification tasks and reduced the need for low-level design decisions. In contrast, DCL-KDM’s generic abstractions required participants to choose among multiple types (e.g., subsystem, layer, component), increasing decision effort and introducing variability.  

\begin{figure}[ht]
	\centering
	\includegraphics[width=0.7\textwidth]{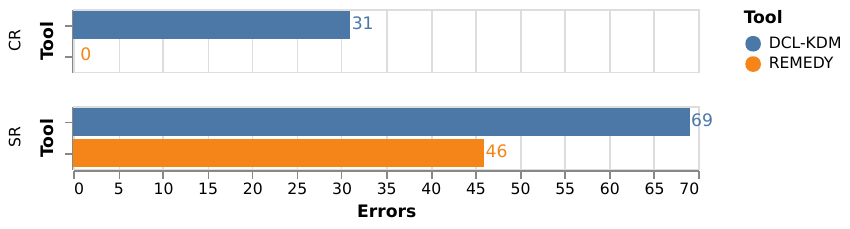}
	\caption{Errors observed with both tools}\label{fig:bothtools}
	\vspace{-2mm}
\end{figure}

Error analysis further supports this interpretation. As shown in Figure~\ref{fig:bothtools}, participants committed $115$ errors with DCL-KDM, $46$ related to structural rules (SR) and $69$ to communication rules (CR). In contrast, only $31$ errors occurred with REMEDY, all in communication rules. The higher error rate in DCL-KDM can be attributed to its flexible but less constrained abstraction hierarchy, where subsystems and layers can be deeply nested, complicating specification and increasing the likelihood of structural mistakes. In addition, DCL-KDM lacked built-in validation for duplicate or circular rules, both of which were observed in the experiment.  

Errors in REMEDY, though fewer, highlighted a specific limitation: most were associated with low-level communication rules across control loops. These errors often resulted from copy–paste operations involving reference inputs of different loops. Indeed, this revealed a bug in REMEDY's validation mechanism, where access rules between loops were not enforced for certain low-level abstractions. This issue was identified through the experiment and corrected in subsequent tool versions.  

Effort ratings also highlight the difference in user experience. Participants consistently rated REMEDY as easier to use, with a mean score of $1.70$ (close to “easy”), while DCL-KDM averaged $2.83$ (closer to \say{difficult}). These perceptions align with the statistical results and further reinforce the benefit of domain-specific support.  

The analysis confirms that domain-specific approaches such as REMEDY improve productivity not only by reducing specification time but also by lowering error rates and user effort. While DCL-KDM provides flexibility as a generic language, its lack of domain focus introduces ambiguity and complexity, leading to reduced efficiency and greater likelihood of mistakes. The experiment thus provides empirical validation for the claim that domain-specific ACC approaches can enhance both the effectiveness and usability of architectural conformance checking in self-adaptive systems.

\subsection{Threats of Validity} \label{sec:threats}

This section presents the limitations and threats to validity of our study. While we took several steps to reduce risks, certain threats remain that could affect the interpretation and generalization of our findings.

\subsubsection{Threats to Internal Validity}

Internal validity refers to the extent to which the observed effects on the dependent variables can be attributed to the independent variables. Due to the SARS-CoV-2 pandemic, the controlled experiment was conducted online. As a result, the instructor could not directly verify whether participants interacted with each other regarding the architectural specifications.  

To reduce this threat, participants were informed before the experiment that it was not a competition and that there were no rewards for finishing earlier or producing a more optimal specification. After analyzing both the pilot and the experimental results, we observed heterogeneity in the submitted specifications. This suggests that even if some interaction among participants occurred, it did not significantly influence the outcomes of the experiment.  

Another internal validity concern is the relatively small sample size of participants, which may reduce statistical power and limit the robustness of the conclusions. Future replications with larger and more diverse samples are needed to strengthen the findings.

\subsubsection{Threats to Construct Validity}

Construct validity concerns the degree to which the measures used in the study accurately represent the intended constructs. A potential threat here was the inadequate pre-operational definition of constructs. Although all relevant concepts were introduced to participants, the depth of explanation might not have been sufficient for some, given the restrictions on the number and duration of online sessions.  

To mitigate this threat, we collected participant profiles and compared them with their academic records to ensure alignment between the required competencies and the delivered material. Furthermore, one of the researchers involved in the study was responsible for teaching at least one course that participants had attended. This provided additional confidence in the baseline knowledge of the participants.  

In addition, productivity and error rate were used as proxies for modeling effort and precision. While appropriate, these metrics do not capture broader architectural quality attributes, such as maintainability, scalability, or long-term evolution.

\subsubsection{Threats to External Validity and Conclusion Validity}

External validity refers to the extent to which the findings can be generalized to other populations and contexts, while conclusion validity concerns the soundness of the inferences drawn from the results. Our study involved undergraduate students in Computer Engineering. Although they are not professional software engineers, they represent the upcoming generation of practitioners \citep{Kitchenham2002}. A further threat lies in the possibility that the architectural specifications used in the study may not directly correspond to those in real-world applications. To reduce this risk, the specifications were designed based on architectural patterns reported in influential studies of the field.  

Another concern relates to the reliability of time measurements. Participants were asked to record their start and end times, which could introduce errors if they forgot to register them. To mitigate this, we used an online chronometer that participants shared with the instructor, ensuring that the recorded times were consistent and verifiable.  

Beyond these, the evaluation relied on a single case study (a robotic self-adaptive system) and one controlled experiment. While these illustrate the feasibility and benefits of REMEDY, generalizability to industrial-scale SASs remains limited. Replications in other domains are required to confirm the external validity of the approach.

\subsubsection{Additional Tool- and Process-Related Threats}

There are limitations related to tool support and the recovery process. REMEDY was implemented as an Eclipse-based plugin relying on KDM/OCL infrastructure. This may restrict adoption in environments that use other platforms or industrial pipelines.  

Moreover, the recovery of the Current Architecture requires manual intervention and reliance on naming conventions. This introduces potential subjectivity, as different architects may perform mappings inconsistently, which could affect reproducibility and accuracy of results.

\section{Related Works} \label{related}

A variety of approaches for Architecture Conformance Checking (ACC) have been developed to detect architectural drift. Once identified, the objective is to refactor the Current Architecture (CA) to align it with the Planned Architecture (PA). Generic approaches are typically characterized by the use of a general vocabulary to describe architectural abstractions such as entity, layer, module, and subsystem. Some techniques rely directly on source-code concepts (e.g., packages and classes) to denote components or modules, which are then used to specify the PA \citep{Brunet2012,Gurgel2014}.  

One of the earliest approaches for detecting architectural violations is the reflexion model \citep{Holt2000}. This technique compares a high-level model containing entities and their relationships with a low-level model, usually extracted from source code and represented as a call graph. A mapping between these two models must be established, where entities from the high-level model are associated with elements of the source model, often using regular expressions. Based on this mapping, a reflexion model is produced that highlights where the high-level specification conforms, or fails to conform, to the source model \citep{Holt2000,Murphy1995}.  

Tools implementing this approach include the work of \citet{Brunet2012} and \citet{Herold2015}. The first integrates a \textit{Visual diff}\footnote{https://w3.cs.jmu.edu/bernstdh/web/common/tools/diff.php} tool and manual inspection to detect violations related to method calls, field access, generalization, exceptions, return types, and parameter usage. The second introduces \textit{JITTAC}\footnote{http://actool.sourceforge.net/}, which compares models and identifies violations concerning package, class, and method dependencies in Java.  

Another line of research relies on Dependency Structure Matrices (DSM), which represent components in rows and columns, with intersections denoting relations among them \citep{Baldwin2000,Sangal2005}. DSMs provide a compact way to visualize dependencies, support grouping of components into modules, and facilitate hierarchical analysis at different levels of abstraction. DSMs can also be integrated with architectural rules, where constraints are defined to allow or forbid dependencies among entities. A notable implementation is \textit{Lattix Architect}\footnote{http://lattix.com}, which offers a graphical interface for DSMs, enabling architects to define constraints and navigate rule violations or cycles in system dependencies.  

Query-based approaches also exist, such as \textit{CQLink}\footnote{http://www.ndepend.com} and \textit{Semmle .QL}\footnote{https://semmle.com/}, which adopt SQL-inspired syntax to express rules. These tools allow users to query code entities (e.g., types, methods, namespaces) and check them against predefined values. While effective at the source-code level, their focus on low-level abstractions limits their ability to represent high-level architectural concepts, making architectural understanding more challenging.  

ArchLint \citep{Maffort2016} combines static code analysis, change history analysis, and SQL-based heuristics to verify architectural conformance. It requires a version control system and a high-level architectural description. Using these inputs, it detects absences or divergences in dependencies and classifies them accordingly.  

Other approaches leverage dedicated Domain-Specific Languages (DSLs) to define architectural rules. Examples include \textit{DCL 2.0} \citep{Rocha2017}, \textit{DCL-KDM} \citep{Landi2017}, \textit{TamDera} \citep{Gurgel2014}, and \textit{InCode.Rules} \citep{Marinescu2010}. These tools specify constraints on code dependencies such as accesses, declarations, and inheritance.  

\textit{InCode.Rules} supports detection of design flaws (e.g., god classes, data classes) by expressing rules as constraints. DCL 2.0 enables constraints to be written as single statements with a clear syntactical structure. TamDera introduces hierarchical concepts to improve modularity. InCode.Rules further supports rule composition, where rules can define exceptions to others. DCL-KDM extends DCL 2.0 and incorporates support for specifying and serializing PAs based on KDM models, particularly targeting ADM-based projects.  

These DSL-based tools share characteristics of usability and precise specification languages. For example, the authors of DCL 2.0 argue that their language is easier to use than logic-inspired alternatives, which often rely on more complex notation and require familiarity with logic programming \citep{Lozano2015}. Rocha et al. \citep{Rocha2017} compared DCL 2.0 with approaches based on reflexion models and DSMs, noting that it offers greater expressiveness, reusability, and support for a wider variety of constraint types. Being declarative, such DSLs allow rules to be specified without requiring advanced programming skills, using constructs provided by the supported notation.  

Elizondo et al. \citep{Elizondo2017} focused on the Model-View-Controller (MVC) style, characterizing a set of architectural smells such as forbidden layer accesses. They implemented a static analysis tool, \textit{PHP\_CodeSniffer}, to check PHP code against predefined coding standards. The tool tokenizes code, applies textual analysis, and outputs violations with error messages and line numbers. This demonstrates how coding standards can act as architectural rules in practice.  

REMEDY was introduced as a DSL specifically tailored for Self-Adaptive Systems (SASs). Unlike generic ACC tools, it adopts the vocabulary of the MAPE-K reference model, which is widely recognized in the design of adaptive systems. REMEDY enables specification of both PAs and their structural and communication rules using domain concepts. This domain focus allows architects to work with abstractions directly relevant to SASs, producing concise specifications and incorporating domain rules automatically. In doing so, REMEDY reduces the effort required to encode canonical rules manually and supports productivity in architectural conformance checking for adaptive systems.

\section{Conclusion} \label{conclusion}

This paper introduced REMEDY, a domain-specific method for Architectural Conformance Checking (ACC) in Self-Adaptive Systems (SASs) structured around the MAPE-K reference model. Architectural drift is a persistent challenge in these systems due to the evolving nature of adaptive behavior and the complexity of their feedback loops. REMEDY addresses this problem through an integrated solution that combines specification, recovery, and verification using domain-aware abstractions and processes.

The method comprises three core components: a domain-specific language (DSL-REMEDY) for specifying planned architectures using MAPE-K elements; a model recovery mechanism that maps source code artifacts to their architectural counterparts based on the Knowledge Discovery Metamodel (KDM); and an automated conformance checking engine that identifies violations by executing OCL constraints over the recovered models. Together, these elements enable the early detection of inconsistencies and support architects in maintaining alignment between implementation and design.

Unlike generic ACC approaches, which often impose a higher cognitive load by requiring the manual definition of architectural roles and communication rules, REMEDY capitalizes on known domain structures to reduce modeling effort and error-proneness. Moreover, the DSL includes both canonical MAPE-K abstractions and additional low-level elements commonly found in SASs, offering increased expressiveness and realism in architectural modeling. Its custom validators also provide guidance during the modeling process, contributing to better specification quality.

We evaluated REMEDY through two complementary empirical validations: a case study and a controlled experiment. The case study applied REMEDY to a robotic SAS involving hierarchical feedback loops and demonstrated its applicability to realistic adaptive scenarios. The controlled experiment, conducted with software engineering students, showed that REMEDY significantly reduced modeling time and the number of specification errors compared to a general-purpose DSL. These results support the claim that domain-specific ACC methods like REMEDY can improve productivity and accuracy in architectural specification tasks.

Some limitations remain. REMEDY focuses exclusively on the feedback loop and does not yet model adaptation goals, which are essential for runtime reasoning and decision-making. We chose to exclude these goals to keep the DSL concise and focused, but we acknowledge their importance for a full representation of SAS behavior. In future work, we plan to extend REMEDY to include adaptation goals, support additional MAPE-K elements, and validate the method in broader contexts and domains. We also aim to enhance visualization features and explore the integration of predictive techniques for proactive drift detection. 

Finally, as discussed in Section~\ref{sec:threats}, our evaluation is subject to several threats to validity (e.g., limited subject pool, case study scope, tool dependence, and mapping subjectivity). While these do not invalidate the findings, they highlight the need for replication in industrial contexts and with larger samples. Addressing these threats will be an important step in consolidating REMEDY as a comprehensive solution for maintaining architectural integrity in adaptive software systems.

\backmatter

\section*{Declarations}

\begin{itemize}
\item Funding: The authors did not receive support from any organization for the submitted work.
\item Conflict of interest/Competing interests: The authors have no conflicts of interest to declare that are relevant to the content of this article.
\item Data availability:  \url{https://doi.org/10.5281/zenodo.8175111}
\item Author contribution:
\begin{itemize}
    \item Daniel San Martín: Conceived the study, designed the methodology, developed the REMEDY language and tooling, conducted the case study and experiment, and wrote the initial draft of the manuscript.

\item Valter Vieira de Camargo: Provided foundational input on domain-specific languages and architectural conformance, contributed to the theoretical framing of the approach, and was actively involved in revising and refining all sections of the manuscript.

\item Guisella Angulo: Contributed to the validation of the methodology and the interpretation of results, and performed critical review and feedback to improve the clarity and coherence of the manuscript.
\end{itemize}

\end{itemize}

\bibliography{biblio}

\end{document}